\documentclass[10pt,prd,preprintnumbers,floatfix,aps,nofootinbib,showpacs,twocolumn,superscriptaddress,noshowpacs]{revtex4-1} 
\usepackage[utf8]{inputenc}
\usepackage[dvipsnames]{xcolor}
\usepackage{graphicx}
\usepackage{hyperref}
\usepackage{amsthm,amsmath,amssymb,amsfonts}
\usepackage{amsbsy}
\usepackage{bbm}
\usepackage{mathtools}
\usepackage{microtype}
\usepackage{tikz}

\makeatletter

\pdfpageheight\paperheight
\pdfpagewidth\paperwidth

\@ifundefined{textcolor}{}{%
 \definecolor{BLACK}{gray}{0}
 \definecolor{WHITE}{gray}{1}
 \definecolor{RED}{rgb}{1,0,0}
 \definecolor{GREEN}{rgb}{0,1,0}
 \definecolor{BLUE}{rgb}{0,0,1}
 \definecolor{CYAN}{cmyk}{1,0,0,0}
 \definecolor{MAGENTA}{cmyk}{0,1,0,0}
 \definecolor{YELLOW}{cmyk}{0,0,1,0}
}

\makeatother

\makeatletter

\DeclareRobustCommand{\rcite}[1]{%
  \rcite@aux#1,\@nil{#1}%
}
\def\rcite@aux#1,#2\@nil#3{%
  \if\relax#2\relax
    Ref.~\cite{#3}%
  \else
    Refs.~\cite{#3}%
  \fi
}
\makeatother
\definecolor{wine}{RGB}{136,34,85}
\definecolor{teal}{RGB}{0,85,102}
\hypersetup{
    colorlinks = true,
    citecolor = {wine},
    linkcolor = {teal},
    urlcolor = {teal},
}

\newcommand{\be}{\begin{equation}}
\newcommand{\ee}{\end{equation}}
\newcommand{\ba}{\begin{eqnarray}}
\newcommand{\ea}{\end{eqnarray}}

\newcommand{\dd}{\mathrm{d}}

\begin{document}

\preprint{IFT-UAM/CSIC-26-57}

\title{DESI and Gravitational Wave Constraints Challenge Quintessential $\alpha$-Attractor Inflation}

\author{Changcheng Jing}
\email{changcheng.jing@estudiante.uam.es}
\affiliation{Instituto de F\'isica Te\'orica (IFT) UAM-CSIC, C/ Nicol\'as Cabrera 13-15, Campus de Cantoblanco UAM, 28049 Madrid, Spain}

\author{George Alestas}
\email{g.alestas@csic.es}
\affiliation{Instituto de F\'isica Te\'orica (IFT) UAM-CSIC, C/ Nicol\'as Cabrera 13-15, Campus de Cantoblanco UAM, 28049 Madrid, Spain}

\author{Sachiko Kuroyanagi}
\email{sachiko.kuroyanagi@csic.es}
\affiliation{Instituto de F\'isica Te\'orica (IFT) UAM-CSIC, C/ Nicol\'as Cabrera 13-15, Campus de Cantoblanco UAM, 28049 Madrid, Spain}
\affiliation{Department of Physics and Astrophysics, Nagoya University, Nagoya, 464-8602, Japan}

\date{\today}

\begin{abstract}
Quintessential inflation models provide a framework that simultaneously describes inflation and dynamical dark energy, 
the latter of which has recently received growing support from DESI observations. A distinctive feature of these models is the kination phase after inflation, which enhances primordial gravitational waves at high frequencies. In this work, we study a class of $\alpha$-attractor quintessential inflation models using a fully numerical approach that follows the scalar-field evolution from inflation to the dark-energy-dominated era, allowing us to compute with high precision both the dynamics of dark energy and the primordial gravitational wave spectrum. Using the latest observational data, including DESI and ACT, we constrain the model parameters and show that the model becomes disfavored once constraints from the gravitational-wave contribution to the effective number of relativistic degrees of freedom, $\Delta N_{\rm eff}$, are included. This is because the model predicts a scalar spectral index $n_s$ that becomes too small to remain consistent with observations when the gravitational-wave abundance is constrained to stay below the $\Delta N_{\rm eff}$ bound. Finally, we present the resulting primordial gravitational wave power spectrum computed using our constrained parameter values, which highlights prospects for detection by future CMB $B$-mode experiments at low frequencies and by gravitational-wave interferometer experiments at high frequencies.

\end{abstract}

\maketitle
%

\section{Introduction \label{sec:introduction}}
The observed accelerated expansion of the Universe at late times remains one of the most profound mysteries in modern cosmology.  Over the past two decades, $\Lambda$CDM model has 
served as the standard cosmological framework, achieving
remarkable success in explaining a wide range of observational data, including Type Ia Supernova (SnIa), the Cosmic Microwave Background (CMB), and large-scale structure (LSS) \cite{Planck:2018vyg,Perlmutter_1999,Riess_1998}.
Despite its successes, the $\Lambda$CDM model exhibits several growing internal inconsistencies among different cosmological probes, particularly between early and late time measurements of the expansion history and structure growth \cite{Perivolaropoulos:2021jda,Riess:2021jrx,Heymans_2021}. Recently, Baryon Acoustic Oscillations (BAO) measurements from the Dark Energy Spectroscopic Instrument (DESI) 
have provided hints consistent with
dynamical dark energy \cite{DESI:2024mwx, DESI2025_BAO_DR2}. 

The late-time accelerated expansion is not the only epoch of accelerated expansion indicated by observations. An earlier period of accelerated expansion,
known as inflation, is required in the very early Universe to solve the horizon, flatness, and monopole problems, as well as to generate the primordial density perturbations that seed cosmic structure formation \cite{sasaki_large_1986,LINDE1982389,PhysRevD.23.347,LINDE1983177}. 
This scenario is strongly supported by observations of the CMB and LSS.
In the standard inflationary paradigm, inflation is driven by a scalar field that satisfies the slow-roll conditions during the inflationary epoch \cite{PhysRevD.50.7222,peebles_quintessential_1999}. While the scalar field rolls slowly along a sufficiently flat region of its potential, the Universe undergoes accelerated expansion. Inflation ends when the field leaves this flat region, after which it typically oscillates around the minimum of its potential. These oscillations initiate the reheating process, during which the inflaton's energy is transferred to radiation, thereby populating the Universe with Standard Model particles \cite{kofman_towards_1997}.

The 
quintessential inflation model provides a unified framework for 
explaining both inflation and dynamical dark energy with a single scalar field
\cite{AresteSalo:2021wgb, de_Haro_2021}. In this scenario, inflation does not end with an oscillatory reheating phase as in conventional models. Instead, after inflation, the inflaton rolls through a steep region of the potential and enters another asymptotically flat region with a much lower energy scale. Consequently, the inflaton survives after inflation and later acts as a quintessence field, driving the late-time accelerated expansion of the Universe. 

Due to the absence of an oscillatory phase, conventional reheating mechanisms do not operate in 
quintessential inflation. As a result, alternative reheating mechanisms that do not rely on inflaton oscillations are required, such as gravitational reheating \cite{Chun:2009yu,Haque:2022kez}, instant preheating \cite{PhysRevD.59.123523,PhysRevD.97.063525}, and curvaton reheating \cite{feng_curvaton_2003}. A minimal level of reheating via gravitational particle production is generically expected, although its efficiency is highly model-dependent and can be negligible for conformally coupled species.  

The Universe undergoes a so-called kination (or stiff) epoch shortly after inflation, during which the kinetic energy of the inflaton dominates the total energy density. Since 
the energy density of the kination component decays as $a^{-6}$, while the radiation energy density evolves as $a^{-4}$,
radiation becomes dominant and the Universe naturally transitions into the standard radiation-dominated era.
The duration of the kination epoch is highly sensitive to the reheating mechanism: if efficient reheating processes operate, the kination phase is short-lived, whereas if gravitational reheating is the sole mechanism, the kination phase can persist for a very long period.
Primordial gravitational wave (GW) modes that reenter the horizon during this epoch experience a significant enhancement in their 
power spectrum 
at high frequencies \cite{Giovannini:1998bp,Peebles:1998qn,Giovannini:1999bh,Giovannini:1999qj,Giovannini:2008tm,Figueroa:2019paj,Duval:2024jsg}. 
Overproduction of primordial GWs can violate Big Bang Nucleosynthesis (BBN) constraints, and therefore provides a powerful way to constrain inefficient reheating scenarios with low reheating temperatures\cite{Giovannini:1998bp,Tashiro:2003qp,Figueroa:2018twl,Haque:2022kez}.

In this paper, we focus on $\alpha$-attractor quintessential inflation~\cite{Akrami:2017cir,Akrami_2021}, which provides a unified description of early- and late-time cosmic acceleration. The early inflationary phase can be rigorously tested by the CMB through observations of primordial density fluctuations \cite{Bhattacharya:2022akq,Kallosh:2025rni,Kallosh:2025ijd,Mondal:2025kur,Maity:2025czp,Wolf:2025ecy,Heidarian:2025drk,Zhu:2025twm,Iacconi:2025odq}. At late times, the model predicts a dynamical dark energy component that can be observationally distinguished from the $\Lambda$CDM scenario, through modifications to both the background expansion history and the formation of LSS~\cite{alestas_desi_2025}.

Despite the extensive literature studying observational constraints on this model, the new aspect of this paper is the inclusion of GW constraints, which has been emphasized in previous works~\cite{WaliHossain:2014usl,Dimopoulos:2017zvq,Wang:2024sgo,alestas_desi_2025,Iacconi:2025odq}, but is explored here in significantly greater detail and with higher precision. The enhancement of high-frequency GWs during the kination epoch contributes an additional amount to the effective relativistic energy density at the time of recombination. This extra radiation component is tightly constrained by recent ACT measurements of the additional contribution to the effective number of relativistic species, $\Delta N_{\rm eff}$~\cite{AtacamaCosmologyTelescope:2025nti,Goldstein:2026iuu}. 

To evaluate the GW contribution with sufficient accuracy, we employ a numerical approach that tracks the evolution of the scalar field from inflation, through reheating, and into the late-time dark-energy-dominated epoch. Such a numerical treatment is essential for correctly estimating the GW amplitude at high frequencies. The contribution of primordial GWs to $\Delta N_{\rm eff}$ is dominated by highest-frequency modes corresponding to the Hubble scale at the end of inflation. Although the GW spectrum is often approximated by imposing an abrupt cutoff at this scale, the true spectral shape should instead reflect the smooth transition of the Hubble parameter from inflation into the kination phase. Moreover, at these high frequencies, associated with modes that exited the horizon near the end of inflation, the slow-roll approximation breaks down~\cite{Kuroyanagi:2008ye, Kuroyanagi:2011iw,PhysRevD.110.103529}. A fully numerical evolution is therefore required to obtain a reliable estimate of GW amplitude. 

In this paper, we first numerically compute the primordial GW power spectrum and then evaluate the resulting GW contribution to $\Delta N_{\rm eff}$. We then use the latest CMB, BAO, and SnIa data to constrain the $\alpha$-attractor quintessential inflation model and compute its Bayesian evidence in comparison with the $\Lambda$CDM model. Our analysis incorporates the newest DESI~\cite{DESI:2024lzq} and ACT~\cite{AtacamaCosmologyTelescope:2025blo} data, which offer important insights into the evolution of dark energy as well as constraints on the effective number of relativistic species, $\Delta N_{\rm eff}$.

Our paper is organized as follows. In Sec. ~\ref{sec:alpha_model}, we provide an overview of the $\alpha$-attractor quintessential inflation model and instant preheating, and build our dynamic equations of the Universe. 
In Sec. ~\ref {sec:GWs}, we provide an overview of primordial GWs, compute the GW power spectrum for the $\alpha$-attractor quintessential inflation model, and discuss the constraints from BBN and CMB, as well as the relationship between GWs and dark energy.
In Sec. ~\ref {sec:MCMC data}, we describe the details of our Markov Chain Monte Carlo (MCMC) analysis and the cosmological data sets used in this work.
In Sec. ~\ref{sec:MCMC result}, we present and discuss our MCMC result.
Finally, we conclude in Sec. ~\ref{sec:conclusions}. 
We use natural units, $ \hbar = c = 1$, and $M_{\rm Pl}=1/\sqrt{8\pi G}$ denotes the reduced Planck mass.

\section{$\alpha$-attractor quintessential inflation model\label{sec:alpha_model}}
We consider the action given by
\begin{equation}
\label{eq:action}
S = \int \dd^4x\sqrt{-g}\left[\frac{1}{2}M_{\rm Pl}^2R - \frac{\partial_\mu\phi\partial^\mu\phi}{2\left(1 - \frac{\phi^2}{6\alpha}\right)^2} - V(\phi) + \mathcal{L}_{\text{m}} \right],
\end{equation}
where 
$R$ is the Ricci scalar in four spacetime dimensions, $V(\phi)$ represents the potential associated with the scalar field $\phi(\vec{x}, t)$, and $\mathcal{L}_\text{m}$ corresponds to the matter Lagrangian density, encompassing the degrees of freedom of the Standard Model. 
The parameter $\alpha$ controls the curvature of field space in the $\alpha$-attractor models and is typically $\mathcal{O}(1)$. Note that, throughout the paper, we rescale the field $\phi$ by $M_{\rm Pl}$ and $\alpha$ by $M_{\rm Pl}^2$.

For the potential, following Refs.~\cite{Akrami_2021,Akrami:2017cir,Dimopoulos:2017zvq}, we consider the form
\begin{equation}
\label{eq: potential_phi}
V(\phi) = M^2e^{\gamma\left(\frac{\phi}{\sqrt{6\alpha}}-1\right)} + V_0,
\end{equation}
which is naturally motivated by high-energy theories such as supergravity or string theory.

There are two poles in the original Lagrangian at $\phi=\pm\sqrt{6\alpha}$.
To get a canonical kinetic term in the action, we use the field redefinition
\begin{equation}
\label{eq:canonical_transformation_phi}
\phi = \sqrt{6\alpha} \tanh\frac{\varphi}{\sqrt{6\alpha}}.
\end{equation}
This maps the poles to $\varphi=\pm\infty$, and the potential becomes
\begin{align}
 \label{eq:pot_varphi}
 V(\varphi) = M^2e^{\gamma\left(\tanh\frac{\varphi}{\sqrt{6\alpha}}-1\right)} + V_0. 
\end{align}
Here, $M^2$, $\alpha$, and $\gamma$ are the parameters characterizing the potential, and $V_0$ is a small constant compared with $M^2$.  
Although models with $3\alpha\in{1,2,3,4,5,6,7}$ are motivated by supergravity and supersymmetry \cite{Ferrara:2016fwe,Kallosh:2017ced,Kallosh:2017wnt}, we treat $\alpha$ as a continuous parameter rather than restricting it to those discrete values so it can be directly constrained by observations as continuous parameters are straightforward to sample with MCMC. The parameter $\gamma$ is not treated as a free parameter in our setup. In quintessential inflation, the field $\varphi$ must source late-time dark energy, so $\gamma$ is determined by a shooting procedure that enforces the required late-time behavior.

\subsection{Inflation}
In the inflation epoch, the effective potential of the inflaton can be approximated by \cite{Dimopoulos_2017}
\begin{align}
    V(\varphi)\simeq M^2\exp\bigg(-2\gamma e^{-\frac{2\varphi}{\sqrt{6\alpha}}}\bigg).
\end{align}
During the slow-roll regime, where the potential is flat,
the exponential term is close to $1$, so $V\sim M^2$.
With this approximation, the slow-roll parameters are 
\begin{align}
\epsilon &:= \frac{M_{\rm Pl}^2}{2}\left(\frac{{\rm d}V/{\rm d}\varphi}{V}\right)^2 \approx \frac{4 \gamma ^2 }{3 \alpha }e^{-\frac{4 \varphi }{\sqrt{6 \alpha }}},\label{Eq:epsilon}\\
\eta &:= M_{\rm Pl}^2\frac{{\rm d}^2V/{\rm d}\varphi^2}{V} \approx -\frac{4 \gamma }{3 \alpha }e^{-\frac{2 \varphi }{\sqrt{6 \alpha }}} \left(1-2 \gamma  e^{-\frac{2 \varphi }{\sqrt{6 \alpha }}}\right). 
\end{align}
The tensor-to-scalar index ratio and the spectral index of scalar curvature perturbations can be written in terms of the e-folding number of inflation evaluated at the CMB pivot scale $k_*$ and denoted by $N_{*}$, as 
\begin{align}
\label{eq:n_s_alpha}
&n_s=1-\frac{2}{N_{*}},\\
\label{eq:r_alpha}
&r=12\alpha\left(N_{*}+\frac{\sqrt{3\alpha}}{2}\right)^{-2}.
\end{align}
The amplitude of primordial scalar perturbations fixes $M$ via
\begin{align}
\label{eq:A_M}
\frac{M^2}{M_{\rm Pl}^4}=\frac{144\pi^2\alpha N_{*}}{2N_{*}-3\alpha}A_s.
\end{align}
Imposing $\epsilon=1$ at the end of inflation yields
\begin{align}
\label{eq:phi_end}
&\varphi_{\rm end}=\frac{1}{2} \sqrt{6 \alpha } \ln \left(\frac{2 \gamma }{\sqrt{3 \alpha }}\right),\\
\label{eq:V_end}
&V_{\rm end}=M^2 e^{-\sqrt{3\alpha }} .
\end{align}
Since $\gamma \sim \mathcal{O}(100)$ when $\alpha \sim \mathcal{O}(1)$~\cite{Akrami_2021}, $2\gamma/\sqrt{3\alpha}$ is always greater than $1$, which implies $\varphi_{\rm end}>0$. Therefore, inflation always ends before the inflaton crosses $\phi=0$, where we assume instantaneous reheating occurs.

\subsection{Preheating}
\label{sec:preheating}
In quintessential inflation, the inflaton does not oscillate at the end of inflation, so the conventional reheating mechanism based on oscillations and parametric resonance does not operate. Instead, one can consider gravitational reheating \cite{Chun:2009yu,Haque:2022kez}, instant preheating \cite{PhysRevD.59.123523,PhysRevD.97.063525}, or curvaton reheating \cite{feng_curvaton_2003}. 

In this work, we focus on instant preheating. 
Following Refs.~\cite{PhysRevD.59.123523,PhysRevD.97.063525}, we consider the following matter Lagrangian density in the preheating epoch 
\begin{equation}
\mathcal{L}_{\text{m}}\supset -\frac{1}{2}g^2\phi^2\chi^2-h\chi\psi\bar{\psi},
\end{equation}
where $g$ and $h$ are coupling constants, $\chi$ is a massless bosonic field directly coupled to the inflaton $\phi$, and $\psi$ is a fermionic field that interacts with $\chi$. In the instant preheating picture, $\chi$ particles are produced through their interaction with the inflaton and subsequently decay into $\psi$, which is assumed to be light and thus behaves effectively as radiation.

The kinetic term of the field $\phi$ is non-canonical, which makes the study of the particle production process challenging. 
However, massless particles $\chi$ are primarily produced around the enhanced symmetry point (ESP), namely $\phi \sim 0$ with $|\phi|\lesssim\Delta\phi\sim (|\dot{\phi}|/g)^{1/2}$.
Normally, we have $\dot{\phi}<\sqrt{\rho_{\rm end}}\lesssim10^{-5}M_{\rm Pl}^2$. If we consider a sufficiently large $g$, the condition $\Delta \phi < \sqrt{6 \alpha}$ is satisfied and we can approximate $\phi \sim \varphi$, so the kinetic term becomes nearly canonical~\cite{Akrami:2017cir}. The number density of $\chi$ particles produced at the ESP is given by~\cite {PhysRevD.59.123523, PhysRevD.97.063525, Kofman:2004}
\begin{equation}
n_{\chi}=\frac{(g\dot{\phi})^{3/2}}{8\pi^3}.
 \end{equation}
After particle production, the inflaton keeps rolling and $\chi$ acquires a mass $m_{\chi}=g|\phi|$ from its interaction with the inflaton. The inflaton then transfers energy to $\chi$ by increasing its mass $m_{\chi}$. The energy density of $\chi$ particles is given by
\begin{equation}
\rho_{\chi}=m_{\chi}n_{\chi}=g|\phi|n_{\chi}.
\end{equation}
The number density of $\chi$ particles, $n_{\chi}$, produced in this process depends on the coupling constant $g$ and on the velocity of $\phi$ at the ESP. The interaction between the inflaton and the massless bosons $\chi$ also introduces a backreaction on the inflaton. The Klein–Gordon equation for $\varphi$, including this backreaction, is given by
\begin{equation}
\label{eq:inflaton}
\ddot{\varphi}+3H\dot{\varphi}+\frac{{\rm d}V}{{\rm d\varphi}}+g n_\chi  {\rm sech}^2\left(\frac{\varphi}{\sqrt{6\alpha}}\right)\frac{|\varphi|}{\varphi}=0,
\end{equation}
where a dot denotes a derivative with respect to cosmic time $t$ and the Hubble expansion rate is defined as $H:=\dot{a}/a$ with $a:=a(t)$ being the scale factor of the Universe.

On the other hand, $\chi$ particles decay into $\psi$ with decay rate \cite{PhysRevD.59.123523}
\begin{equation}
\Gamma=\frac{h^2 m_{\chi}}{8\pi}=\frac{h^2 g|\phi|}{8\pi},
\end{equation}
which depends both on the coupling constant $h$ and the mass of $\chi$.  In this work, we fix $h = 0.01$. This choice ensures that the decay of $\chi$ into lighter particles can be treated perturbatively, while guaranteeing a decay rate sufficiently larger than the Hubble expansion rate, so that the instant preheating mechanism operates efficiently~\cite{PhysRevD.59.123523}. Since the observable predictions are primarily sensitive to the duration of the kination phase, which scales as $\propto 1/\Gamma$ and depends on a combination of the couplings $g$ and $h$, this degeneracy between $g$ and $h$ cannot be disentangled observationally. For this reason, we fix the value of $h$ in our analysis.\footnote{However, backreaction effects, which are discussed later in this subsection, introduce a very mild dependence that can eventually lead to small differences in the dynamical dark energy evolution. As a result, this slightly lifts the degeneracy between $g$ and $h$.  } 

By comparing the decay timescale $1/\Gamma$ with the evolution timescale of $\phi$, $\phi/\dot{\phi}$, one finds that $\chi$ begins to decay efficiently into $\psi$ when $\phi$ reaches $|\phi| \sim (\dot{\phi}/g)^{1/2}/h \sim \Delta\phi/h$, which is much larger than $\Delta\phi$.  
Consequently, the energy that $\chi$ carries at the moment of production (smaller than $g n_{\chi}\Delta\phi$) is much smaller than the energy it acquires after the $\chi$ production process. We therefore treat all particles produced at $\phi=0$ as initially massless. This differs from the picture in Ref.~\cite{PhysRevD.97.063525}, where $\chi$ decays quickly and its initial energy is more important.

Combining the equations of motion for the scalar field with the Friedmann equation, and introducing the following dimensionless variables,
\begin{equation}
x_{\varphi}:=\frac{\dot{\varphi}}{\sqrt{6}H},\
y_{\varphi}:=\frac{V}{3H^2},\
z:=\frac{\kappa\sqrt{\rho_{\psi}}}{\sqrt{3}H},\
n:= \frac{n_{\chi}}{M_{\rm {Pl}}^3},
\end{equation}
we obtain the dynamical system written in terms of derivatives with respect to the e-folding number $N = \ln a$:
\begin{align}
\label{eq:dynamic equations1}
\frac{{\rm d}x_{\varphi}}{{\rm d}N}& = -\frac{g n}{\sqrt{6} H^2}  {\rm sech}^2 \left(\frac{\varphi}{\sqrt{6\alpha}}\right)\frac{\varphi}{|\varphi|}
- x_{\varphi} - q x_{\varphi}\nonumber\\
& - \sqrt{\frac{3}{2}} \epsilon y_{\varphi},\\
\label{eq:dynamic equations2}
\frac{{\rm d}y_{\varphi}}{{\rm d}N} &= -2 q y_{\varphi} + \left( 4 + \sqrt{6} 
\epsilon x_{\varphi} \right) y_{\varphi}, \\
\label{eq:dynamic equations3}
\frac{{\rm d}\varphi}{{\rm d}N} &= \sqrt{6} x_{\varphi}, \\
\label{eq:dynamic equations4}
\frac{{\rm d}z}{{\rm d}N} &= \frac{ g^2 h^2 \alpha n}{4 \pi h_p^3}   \tanh^2 \left(\frac{\varphi}{\sqrt{6\alpha}}\right)
- 2 q z, \\
\label{eq:dynamic equations5}
\frac{{\rm d}n}{{\rm d}N}  &= - 3 n - \frac{ g h^2 n}{8\pi h_p}\sqrt{6\alpha}   \tanh \left(\frac{|\varphi|}{\sqrt{6\alpha}}\right) \nonumber\\
&+ \frac{ 6^{5/4}}{8\pi^3} |x_\varphi|^{5/2} (g  h_p)^{3/2} \delta(\varphi), \\
\label{eq:dynamic equations6}
\frac{{\rm d}h_p}{{\rm d}N}  &= (q-2)h_p,
\end{align}
where $\rho_{\psi}$ is the energy density of the relativistic particles $\psi$, $\delta(\varphi)$ is the Dirac delta function ensuring that $\chi$ particles are produced only at $\varphi=0$, $\epsilon$ is the slow-roll parameter defined in Eq.~\eqref{Eq:epsilon} but evaluated numerically, and $q := R/(6H^2) = \tfrac{1 - 3 x_\varphi^2 + 3 y_\varphi - z}{2}$. The quantity $h_p$ is the Hubble parameter in units of $M_{\rm Pl}$.

After preheating, the scalar field continues to roll until it loses its kinetic energy and freezes due to Hubble friction. 
We denote the corresponding field value by $\varphi_{\rm f}$.
In this epoch, Eq.~\eqref{eq:inflaton} is dominated by the first two terms, and we obtain
\begin{align}
\label{eq:phifrozzen}
\varphi_{\rm f}=\varphi_{\rm pre}-\sqrt{6}\, {\rm arcsinh}\left(\sqrt{\frac{\Omega_{k,{\rm pre}}}{\Omega_{r,{\rm pre}}}}\right),
\end{align}
where 
$\Omega_{k,{\rm pre}}$ and $\Omega_{r,{\rm pre}}$ denote the fractional energy densities of the kinetic and radiation components, respectively, and the subscript ``pre'' indicates quantities evaluated at the end of the preheating process when the field $\chi$ decays. We also note that, in addition to the Hubble friction term, the backreaction (the last term in Eq.~\eqref{eq:inflaton}) affects $\varphi_{\rm f}$ indirectly by modifying $\varphi_{\rm pre}$.

The radiation produced by instant preheating eventually dominates the kinetic energy, and the temperature of the Universe at the kination–radiation equality is defined as the reheating temperature $T_{\rm reh}$. When backreaction effects are negligible, the scalar field freezes due to Hubble friction after this equality. By contrast, if backreaction is significant, the freeze‑out occurs earlier.

We study the evolution of the quintessential scalar field in the presence of instant preheating by solving Eqs.~(\ref{eq:dynamic equations1})-(\ref{eq:dynamic equations6}). Our numerical calculation does not stop at the end of preheating but follows the evolution of the scalar field up to the present epoch, where it acts as dynamical dark energy. The parameter $\gamma$ in the potential is not treated as a free parameter; instead, it is determined by a shooting procedure to ensure that Eq.~(\ref{eq:dynamic equations6}) yields the correct Hubble constant today, i.e., $h_p(N=0)=H_0/M_{\rm Pl}$.

\subsection{Dynamical dark energy}
\label{sec:dynamic-dark-energy}
In the $\alpha$-attractor model, $V_0$ is the main parameter controlling the dark energy sector. Following Refs.~\cite{Akrami_2021,Akrami:2017cir}, we consider two representative cases, $V_0=0$ and $V_0=-M^2 e^{-2\gamma}$, which we refer to as $\mathrm{Exp}$-model I and $\mathrm{Exp}$-model II, respectively. In $\rm Exp$-model I, the potential minimum of $\phi$ is set to zero (when $\phi\to -\infty$); however, since $\phi$ cannot reach the pole $\phi=-\sqrt{6\alpha}$, it leaves a non-zero cosmological constant $\Lambda=M^2 e^{-2\gamma}$. 
In contrast, $\rm Exp$-model II introduces a non-zero $V_0$ to offset the cosmological constant to ensure the minimum of the potential of $\varphi$ is equal to 0. 


Apart from the shape of the potential, the reheating process also affects the evolution of dark energy by altering $\varphi_{\rm f}$. As discussed in Sec.~\ref{sec:preheating}, reheating determines the final freezing point $\varphi_{\rm f}$ through two effects: backreaction and Hubble friction. 
The value of $\varphi_{\rm f}$ influences dynamical dark energy through ${\rm d}V/{\rm d}\varphi$, so that for different $\varphi_{\rm f}$ the scalar field starts to roll at different redshifts.
We note that in our previous paper~\cite{alestas_desi_2025}, we did not consider the detailed reheating process and instead assumed a fixed value of $\varphi_{\rm f}$, which leads to some differences in the results.

\begin{figure}
\includegraphics[width=0.45\textwidth]{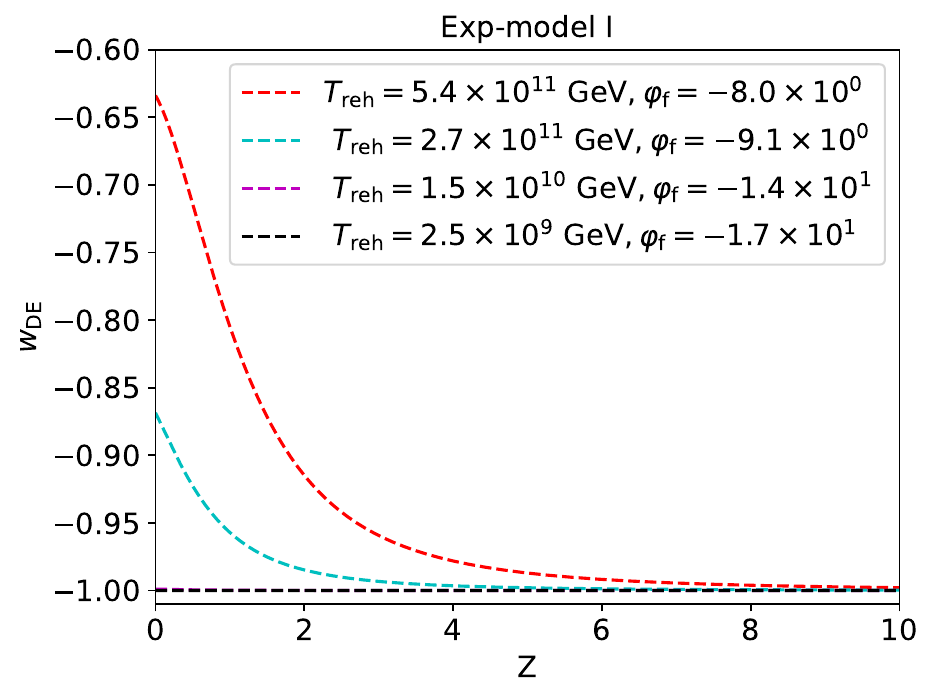}\\  
    \caption{The EOS parameter of dark energy as a function of redshift $z$ for  Exp-model-I. }
    \label{fig:Exp-model-I}
\end{figure}

\begin{figure}
\includegraphics[width=0.45\textwidth]{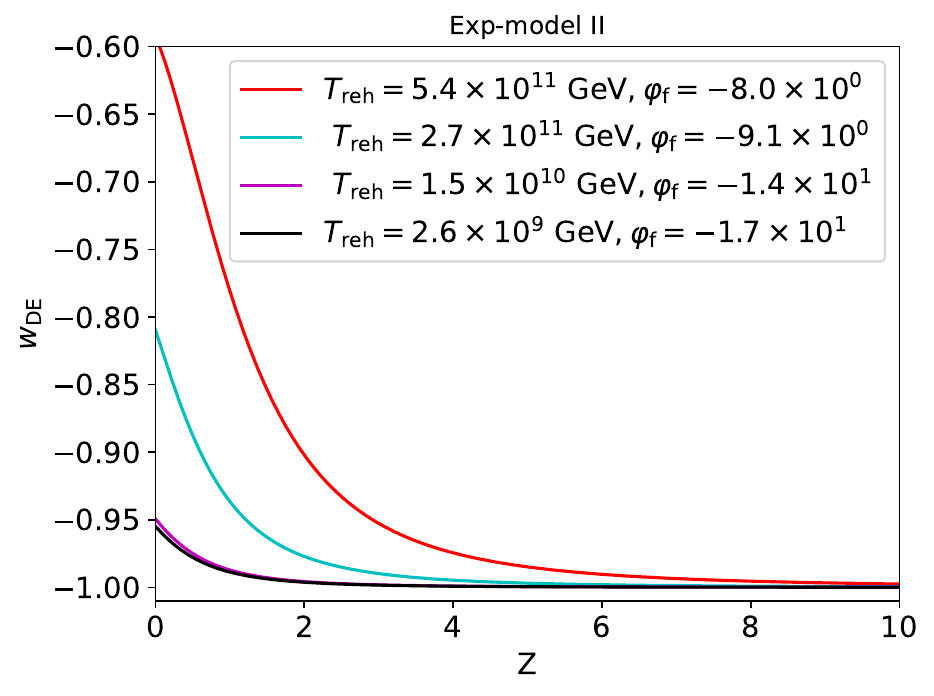}\\  
    \caption{The EOS parameter of dark energy as a function of redshift $z$ for  Exp-model-II. }
    \label{fig:Exp-model-II}
\end{figure}

In Fig.\ref{fig:Exp-model-I} and Fig.\ref{fig:Exp-model-II}, we show the evolution of the dark energy equation-of-state (EOS) parameter $w_{\rm DE}$ as a function of redshift $z$ for Exp-model-I and Exp-model-II, respectively. In both cases we fix $\alpha=7/3$ and $M=2.7\times 10^{-5}M_{\rm Pl}^2$. For Exp-model-I, when the reheating temperature $T_{\rm reh}$ is sufficiently low, the model behaves like $\Lambda$CDM with $w_{\rm DE}=-1$; for higher reheating temperatures it becomes dynamical. For Exp-model-II, $w_{\rm DE}$ is always dynamical, and a larger $T_{\rm reh}$ leads to a more pronounced evolution of $w_{\rm DE}$. Below a certain critical reheating temperature, $T_{\rm reh}$ no longer has a significant impact on the dark energy evolution. These results are in agreement with Ref.~\cite{zhumabek_connecting_2023}.

\section{Stochastic gravitational wave background\label{sec:GWs}}

In this section, we describe the method used to compute the primordial GW spectrum, which originates from quantum fluctuations of tensor modes during inflation. These modes are stretched beyond the horizon and frozen by Hubble friction during inflation. After inflation, they reenter the horizon and propagate as primordial GWs. The Hubble expansion rate at horizon reentry imprints itself on the GW spectrum, so the resulting frequency dependence provides a probe of the early expansion history of the Universe~\cite{Kuroyanagi:2011fy}.

In the standard inflation model, efficient reheating is followed immediately by a radiation-dominated era, producing an almost flat GW spectrum, $\Omega_{\rm GW}(f) \approx 10^{-16}(r/0.01)$, whose tiny amplitude makes it extremely difficult to detect. An intermediate early matter-dominated phase is also possible, but this further suppresses the high-frequency part of the spectrum~\cite{Seto:2003kc}. By contrast, in the $\alpha$-attractor quintessential inflation model, the presence of a kination epoch between inflation and radiation enhances the GW background at high frequencies, potentially bringing it within the sensitivity range of direct detection experiments~\cite{alestas_desi_2025}. These high-frequency GWs also affect CMB and BBN observables as an extra relativistic degree of freedom, which places an upper bound on the GW amplitude and can therefore be translated into a constraint on the duration of the kination phase

\subsection{Primordial gravitational wave spectrum}
Tensor perturbations, satisfying the transverse-traceless (TT) gauge condition and propagating in an FLRW background, are decomposed into Fourier modes. We denote the mode with comoving wavenumber $k$ by $h^{\lambda}_{k}$, where $\lambda$ labels the polarization ($+$ or $\times$).
The equation of motion (EOM) for the GW modes then takes the form
\begin{align}
\label{eq:GW-equation-1}
h_{k}^{''\lambda}(\tau)+2\frac{a^{'} (\tau)}{a(\tau)}
h^{'\lambda}_{k}(\tau)+h^{\lambda}_{k}(\tau)=0,
\end{align} 
where $\tau$ is conformal time defined by $d\tau:=dt/a(t)$, and primes denote derivatives with respect to $\tau$.  
Here, anisotropic stresses, which could alter their evolution at horizon reentry~\cite{Weinberg:2003ur,Watanabe:2006qe,Dent:2013asa,caprini_cosmological_2018}, are neglected as they do not affect the ultra-high-frequency GWs of interest unless there are free-streaming relativistic species beyond the Standard Model~\cite{Jinno:2012xb}. 

Defining $f^{\lambda}_k:= a(\tau) h^{\lambda}_k(\tau)$, Eq.~(\ref{eq:GW-equation-1}) can be rewritten as
\begin{align}
\label{eq:GW-equation-2}
f_{k}^{''\lambda}(\tau)+\omega_{k}^2(\tau)f^{\lambda}_{k}(\tau)=0,
\end{align}
where 
\begin{align}
\label{eq:define-of-omegak}
\omega_{k}^2=k^2-\frac{a''}{a}.
\end{align}
The adiabatic condition $\frac{a''}{a} \ll k^2$ holds both before a mode exits the horizon during inflation and after it re-enters the horizon in the post-inflationary Universe. In these regimes, Eq.~\eqref{eq:GW-equation-2} simplifies to
\begin{align}
\label{eq:GW-equation-2-1}
f^{\prime\prime\lambda}_{k}(\tau) + k^2 f^{\lambda}_{k}(\tau) = 0,
\end{align}
which is equivalent to the equation of motion for a free field in Minkowski spacetime. This allows for a well-defined notion of creation and annihilation operators in both epochs. 

Let $a_k$ and $a_k^\dagger$ denote the annihilation and creation operators before the mode exits the horizon, and $b_k$ and $b_k^\dagger$ the corresponding operators after the mode re-enters. These two sets of operators are related by a Bogoliubov transformation 
\begin{align}
a_k &=  \alpha_{k}^* b_k - \beta_{k}^* b_k^\dagger  \\
b_k &=  \alpha_{k} a_k + \beta_{k} a_k^\dagger .
\end{align}
where $\alpha_{k}$ and $\beta_{k}$ are Bogoliubov coefficients, that match condition
\begin{align}
\label{eq:normalize-condition}
|\alpha_k|^2-|\beta_k|^2=1.
\end{align}

We set the initial condition of Eq.~\eqref{eq:GW-equation-2} for modes in the subhorizon regime during inflation to be the Bunch-Davies vacuum,
\begin{align}
\label{eq:initial-condition}
f^{\lambda}_{k}(\eta \to -\infty) &= \frac{e^{-ik\eta}}{\sqrt{2k}},
\end{align}
Using the Bogoliubov transformation, the solution after the mode re-enters the horizon in the post-inflationary Universe is given by
\begin{align}
\label{eq:end-condition}
f^{\lambda}_{k}(\eta \to +\infty) &= \frac{1}{\sqrt{2k}} \left( \alpha_k e^{-ik\eta} + \beta_k e^{ik\eta} \right),
\end{align}
On the other hand, by solving Eq.~\eqref{eq:GW-equation-2} with the initial condition given in Eq.~\eqref{eq:initial-condition}, we can determine the mode function $f^{\lambda}_{k}$ after the mode has re-entered the horizon. Using Eq.~\eqref{eq:end-condition} together with the normalization condition in Eq.~\eqref{eq:normalize-condition}, the Bogoliubov coefficient $|\beta_k|^2$ can be extracted as
\begin{align}
\label{eq:beta}
|\beta_k|^2 = \frac{k^2 |f^{\lambda}_{k}|^2 + |f_k'|^2 - k}{2k}.
\end{align}

The energy density of GWs generated during inflation can be evaluated directly from the Bogoliubov coefficient $\beta_k$ as~\cite{Chun:2009yu}
\begin{align}
\rho_{\rm GW}=\frac{1}{\pi^2 a^4}\int dk k^3|\beta_k|^2.
\end{align}
The amplitude of the GW background is commonly characterized in terms of the energy-density fraction relative to today’s critical density, $\rho_{c}= 3M_{\rm Pl}^2 H_{0}^2$. Finally, we obtain
\begin{align}
\label{eq:power_spectrum}
\Omega_{\rm GW}(k):= \frac{1}{\rho_{c}}\frac{{\rm d} \rho_{\rm GW}}{{\rm d} \ln{k}}=\frac{k^4}{3\pi^2 M_{\rm Pl}^2 H^2 a^4}|\beta_k|^2.
\end{align}

\subsection{Numerical result}
\label{sec:GW_nu}
To compute the GW spectrum $\Omega_{\rm GW}(k)$, first, we numerically solve Eqs. (\ref{eq:dynamic equations1})-(\ref{eq:dynamic equations6}) to obtain the evolution of the scale factor $a(\tau)$. 
The EOS parameter smoothly transitions from $-1$ (inflation) to $1$ (kination) and then to $1/3$ (radiation), with the detailed behavior determined by both the potential $V(\varphi)$ and the reheating process. Afterward, the Universe enters a matter-dominated epoch, as in standard cosmology, followed by a dark-energy-dominated era driven by the scalar-field dynamics. Using this background solution, we then solve the EOM for GWs, Eq.~(\ref{eq:GW-equation-2}), for each frequency $f=k/(2\pi)$.  Finally, the GW power spectrum $\Omega_{\rm GW}(k)$ is obtained using Eqs.~(\ref{eq:beta}) and (\ref{eq:power_spectrum}). 

\begin{figure}
\includegraphics[width=0.45\textwidth]{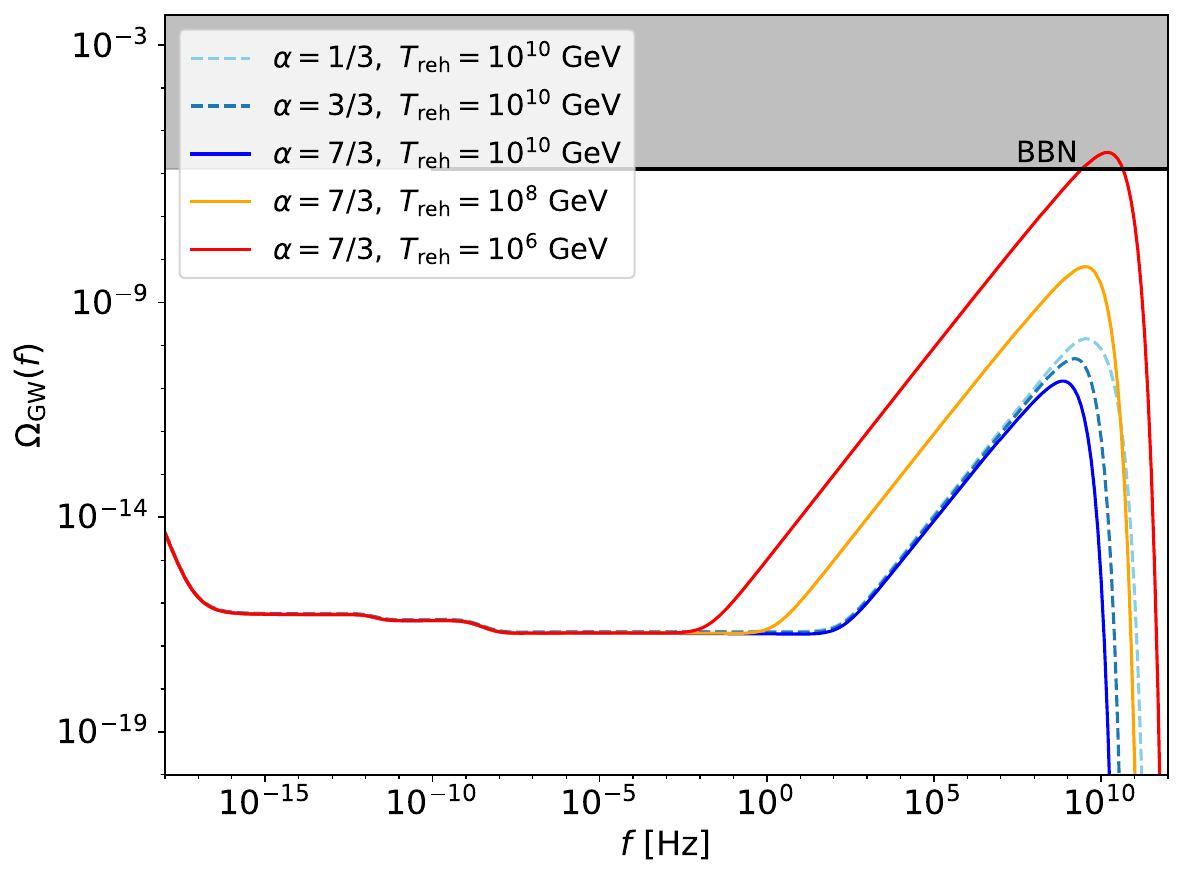}\\  
    \caption{The gravitational power spectrum for different model parameters. For all curves, we fix $M=1.5\times10^{-5}M_{\rm Pl}$. The solid curves correspond to the same value of $\alpha$ but different reheating temperatures $T_{\rm reh}$, while the dashed curves and the blue solid curve have the same $T_{\rm reh}$ but different values of $\alpha$.  }
    \label{fig:GW-spectrum}
\end{figure}

In Fig.~\ref{fig:GW-spectrum}, we show the GW power spectrum for the $\alpha$-attractor inflationary model with $M = 1.5 \times 10^{-5} M_{\rm Pl}^2$. The solid curves correspond to the same value of $\alpha$ but different reheating temperatures $T_{\rm reh}$, taking three representative values: $10^{6}\mathrm{GeV}$ (red), $10^{8}\mathrm{GeV}$ (orange), and $10^{10}\mathrm{GeV}$ (dark blue). The dashed curves and the dark-blue solid curve are computed using the same value of $T_{\rm reh}$ but for three different choices of $\alpha$: $7/3$, $3/3$, and $1/3$. 

As seen in the figure, the GW power spectrum rises steeply at frequencies higher than $f_{\rm reh}$ due to the presence of a kination epoch. This transition frequency corresponds to the mode which reentered the horizon at the time of kination-radiation equality, and is given by~\cite{Figueroa:2019paj} 
\begin{align}
    f_{\rm reh}=3.8\times 10^{-8}\left(\frac{g_{*,\rm{reh}}}{106.75}\right)^{1/2} \left(\frac{g_{*s,\rm{reh}}}{106.75}\right)^{-1/3}\left(\frac{T_{\rm reh}}{\mathrm{GeV}}\right)\mathrm{Hz},
\end{align}
where $g_*$ counts the effective number of relativistic degrees of freedom contributing to the energy density, while $g_{*,s}$ counts those contributing to the entropy density. The subscript “reh” indicates evaluation at the time of kination-radiation equality. By comparing the different solid curves, we see that a lower $T_{\rm reh}$ corresponds to a lower critical frequency $f_{\rm reh}$ and produces a more pronounced peak at high frequencies. This is because a lower reheating temperature implies a longer kination epoch for a fixed inflation scale (i.e., for fixed $M$). As a result, the high-frequency modes spend more time in the kination phase, during which the GW energy density redshifts more slowly than the background energy density. Consequently, $\Omega_{\rm GW}$ becomes enhanced at the highest frequencies. In the figure, we also plot the BBN upper bound, $\Omega_{\rm GW} h^2 < 1.3 \times 10^{-6}$ for $f > 2 \times 10^{-11}\mathrm{Hz}$~\cite{Yeh:2022heq}, which shows that the low-reheating-temperature scenario with $T_{\rm reh} \lesssim 10^{6}\mathrm{GeV}$ would be excluded. A similar constraint will be examined in detail in the next subsection using the CMB bound on $\Delta N_{\rm eff}$.

For GW modes generated near the end of inflation, certain high-frequency modes may never undergo a freeze-out phase. During this stage, the Bogoliubov coefficient $\beta_k$ decreases rapidly, which leads to an ultraviolet cutoff in the GW spectrum at a characteristic frequency $f_{\rm end}$. This cutoff frequency is determined by the Hubble scale at the end of inflation $H_{\rm end}$ and is given by
\begin{align}
    f_{\rm end}&=1.7\times 10^{9} \left(\frac{g_{*,\rm{reh}}}{106.75}\right)^{5/12} \left(\frac{g_{*s,\rm{reh}}}{106.75}\right)^{-1/3}\nonumber\\
    &\quad\times \left(\frac{ 10^{10}{\rm GeV}}{T_{\rm reh}}\right)^{1/3}\left(\frac{H_{\rm end}}{\mathrm{10^{13}GeV}}\right)\mathrm{Hz}.
\end{align}
Using Eq.~\eqref{eq:V_end}, we obtain $f_{\rm end} \propto T_{\rm reh}^{-1/3}H_{\rm end} \propto T_{\rm reh}^{-1/3}\sqrt{V_{\rm end}} = T_{\rm reh}^{-1/3}M \exp(-\sqrt{3\alpha}/2)$. Therefore, models with smaller $\alpha$ and lower $T_{\rm reh}$ predict higher values of $f_{\rm end}$. This trend is consistent with our numerical results, as illustrated in Fig.~\ref{fig:GW-spectrum}.

\subsection{$\Delta N_{\rm eff}$ and $n_s$}

\begin{figure}[h!]
    \includegraphics[width=0.45\textwidth]{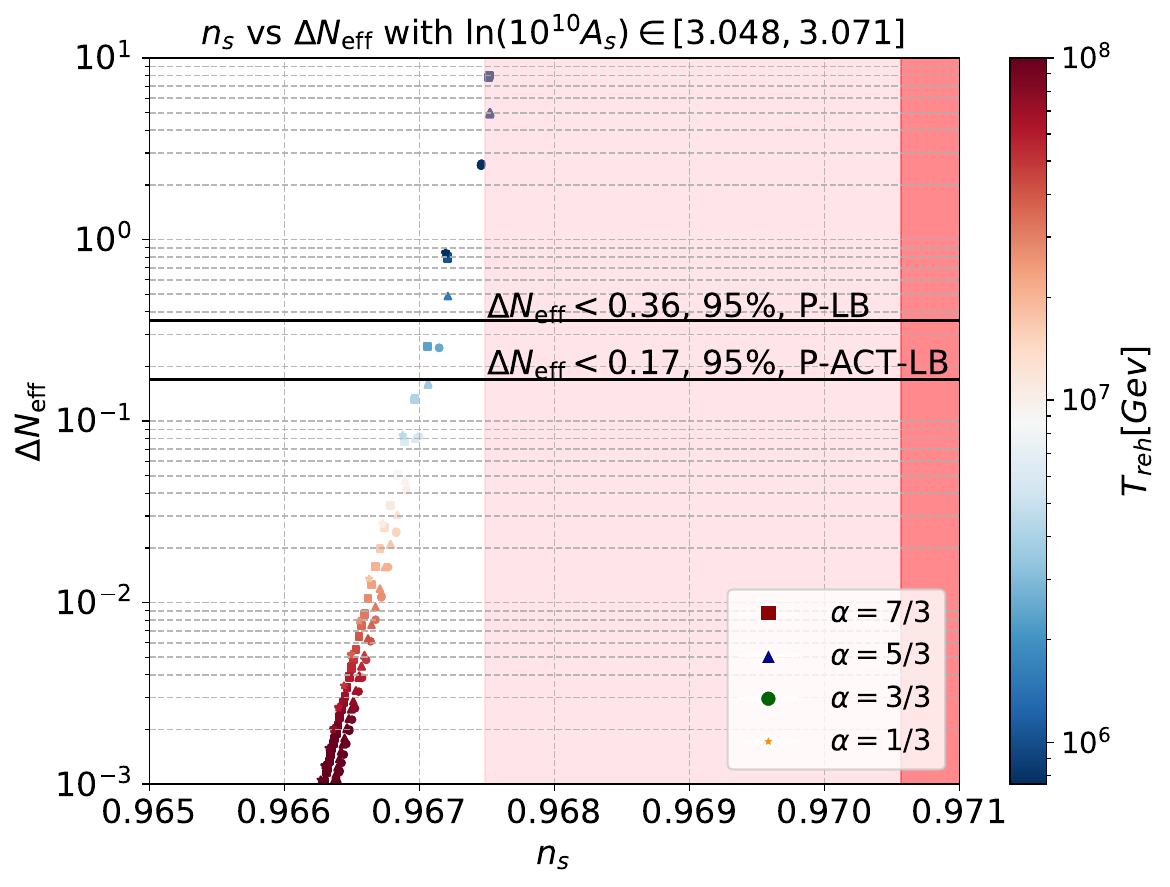}\\  
    \caption{The relationship between the scalar spectral index $n_s$ and the GW-induced modification of the effective number of relativistic species $\Delta N_{\rm eff}$ in the $\alpha$-attractor inflationary model is shown. Different colors of the points correspond to different values of $T_{\rm reh}$, while different shapes correspond to different values of $\alpha$. The pink and red shaded regions denote the $2\sigma$ and $1\sigma$ confidence intervals of $n_s$, respectively, obtained from the $\Lambda$CDM constraints based on P-ACT-LB~\cite{AtacamaCosmologyTelescope:2025blo}. The black solid lines show constraints on $\Delta N_{\rm eff}$~\cite{AtacamaCosmologyTelescope:2025nti}. }
    \label{fig:ns_deltaneff}
\end{figure}

For the $\alpha$-attractor inflation model, the scalar spectral index $n_s$ has a simple relation to the e-folding number $N_*$, as given in Eq.~\eqref{eq:n_s_alpha}. The value of $N_*$ is also linked to the post-inflationary evolution of the Universe and exhibits a mild dependence on the reheating temperature, given by
\begin{align}
\label{eq:N*}
N_{*}&=67-\ln\bigg(\frac{k_*}{H_0 a_0}\bigg)+\frac{1}{4}\ln\bigg(\frac{V_*^2}{\rho_{\rm end}M_{\rm Pl}^4}\bigg)\nonumber\\
-&\frac{1}{12}\ln g_{*,{\rm reh}}+\frac{1-3w_{\rm reh}}{12(1+w_{\rm reh})}\ln\bigg(\frac{\rho_{\rm reh}}{\rho_{\rm end}}\bigg),
\end{align}
where we take the CMB pivot scale to be $k_* = 0.05 {\rm Mpc^{-1}}$. Here, $w_{\rm reh}$ is the EOS parameter during reheating, for which we set $w_{\rm reh}=1$, corresponding to a kination phase; $\rho_{\rm reh}$ is the energy density at kination-radiation equality; and $\rho_{\rm end}$ is the energy density at the end of inflation. 
Through this equation, the observed value of $n_s$ has implications for the reheating scenario.

Moreover, because the duration of the kination epoch determines the enhancement of the high-frequency GW spectrum, the value of $\rho_{\rm reh}$ inferred from $n_s$ can be translated into the amount of extra radiation degrees of freedom, $\Delta N_{\rm eff}$. The energy density of GWs behaves as an additional radiative component in the background energy budget of the Universe. Consequently, their contribution to the total radiation content of the Universe can be encoded in an effective number of neutrino species as \cite{caprini_cosmological_2018}
\begin{align}
\label{eq:definenu}
\int_{f=0}^{f=\infty} d(\log f)\, h_0^2 \Omega_{\mathrm{GW}}(f)= 5.6 \times 10^{-6} \Delta N_{\rm eff},
\end{align}
where we define $\Delta N_{\rm eff}:=N_{\nu}-3.044$ and assume the Standard Model value $N_\nu = 3.044$ for the effective number of neutrino species in the absence of any additional radiation. In this work, we treat the GW background as the only source of extra radiation, contributing an amount of $\Delta N_{\rm eff}$.


The ACT collaboration has obtained a stringent upper bound on the spectral index $n_s=0.974\pm 0.003$~\cite{AtacamaCosmologyTelescope:2025blo} as well as on the amount of extra radiation, $\Delta N_{\rm eff} < 0.17$~\cite{AtacamaCosmologyTelescope:2025nti}. In Fig.~\ref{fig:ns_deltaneff}, we compare these observational constraints with the predictions of the $\alpha$-attractor model. Different colors represent the model predictions for different values of $T_{\rm reh}$ (with decreasing $T_{\rm reh}$ leading to a longer kination epoch and consequently a higher peak amplitude of the primordial GW spectrum), while points with different shapes correspond to different values of $\alpha$. We also show the $1\sigma$ and $2\sigma$ regions of $n_s$, together with the constraint on $\Delta N_{\rm eff}$, obtained from the combined analysis of Planck, ACT, ACT lensing, and DESI DR1 (P-ACT-LB)~\cite{AtacamaCosmologyTelescope:2025blo,AtacamaCosmologyTelescope:2025nti}. The relation between $n_s$ and $\Delta N_{\rm eff}$ is obtained fully numerically by computing the scalar-field evolution, together with the evaluation of the GW power spectrum and the use of Eq.~\eqref{eq:definenu}. To generate the figure, we impose that $\ln(10^{10}A_s)$ for all $\alpha$-attractor models shown lies within the $1\sigma$ confidence region of the P-ACT-LB constraints, so that the mild dependence on the inflation scale $M$ in Eq.~\eqref{eq:N*} is tightly suppressed by the precise measurement of $A_s$.

Recent studies have shown that the CMB data from ACT combined with Planck prefer a higher scalar spectral index than the value inferred from Planck alone. As a result, $\alpha$-attractor models are no longer favored by the combined dataset~\cite{Kallosh:2025rni,Iacconi:2025odq}. Lowering the reheating temperature $T_{\rm reh}$ can help alleviate this tension, since a smaller $T_{\rm reh}$ increases $n_s$. However, this figure highlights an important point: the extent to which $T_{\rm reh}$ can be reduced is strongly constrained by the bound on $\Delta N_{\rm eff}$, which limits the excess GW energy density. Consequently, even the largest value of $n_s \sim 0.967$ allowed for $T_{\rm reh} \sim 10^7{\rm GeV}$ still lies outside the $2\sigma$ region of the P-ACT-LB constraints.

\subsection{Connect GWs with dynamical dark energy}
To realize the dynamic evolution of dark energy in the late Universe, we should let the scalar field roll when the dark energy starts to dominate the Universe. This requirement imposes a condition on the potential of $\alpha$-attractor inflationary models,
\begin{align}
\label{eq:req_dde0}
\frac{|dV(\varphi)/d\varphi|}{V(\varphi)}\bigg |_{\varphi=\varphi_{\rm f}}\geq\mathcal{C}\sim \mathcal{O}(1).
\end{align}
If $\varphi$ is small enough, we can rewrite Eq.~\eqref{eq:req_dde0} as
\begin{align}
\label{eq:req_dde}
\frac{|dV(\varphi)/d\varphi|}{V(\varphi)}\bigg |_{\varphi=\varphi_{\rm f}}\simeq\frac{\sqrt{\frac{2}{3}} e^{\frac{\sqrt{\frac{2}{3}} \varphi_{\rm f} }{\alpha }}}{\alpha  \left(e^{\frac{\sqrt{\frac{2}{3}} \varphi_{\rm f} }{\alpha }}+\frac{\alpha  e^{2 \gamma }}{M^2}V_0+1\right)}\geq\mathcal{C}
\end{align}

For Exp-model I, Eq.~\eqref{eq:req_dde} leads to a lower bound on $\varphi_{\rm f}$,
\begin{align}
\label{eq:phi_f_lower_bound}
\varphi_{\rm f}\geq\sqrt{\frac{3}{2}} \alpha  \log \left(-\frac{3 \alpha  \mathcal{C}}{3 \alpha  \mathcal{C}-\sqrt{6}}\right).
\end{align}
As shown in Eq.~\eqref{eq:phifrozzen},
the value of $\varphi_{\rm f}$ depends on the ratio $\Omega_{\rm k}/\Omega_{\rm r}$. 
To satisfy the condition in Eq.~\eqref{eq:phi_f_lower_bound}, a sufficiently large reheating temperature $T_{\rm reh}$ is required so that enough radiation is produced. This leads to a smaller ratio $\Omega_{\rm k}/\Omega_{\rm r}$, a shorter kination epoch, and consequently a weaker GW signal.

For Exp-model II, Eq.~\eqref{eq:req_dde} leads to
\begin{align}
\alpha<\sqrt{\frac{2}{3\mathcal{C}^2}}.
\end{align}
This condition simply requires $\alpha$ to be smaller than a certain value. Since $\mathcal{C} \sim \mathcal{O}(1)$, the requirement is easily satisfied for $\alpha \sim \mathcal{O}(1)$.

\begin{figure}
\includegraphics[width=0.45\textwidth]{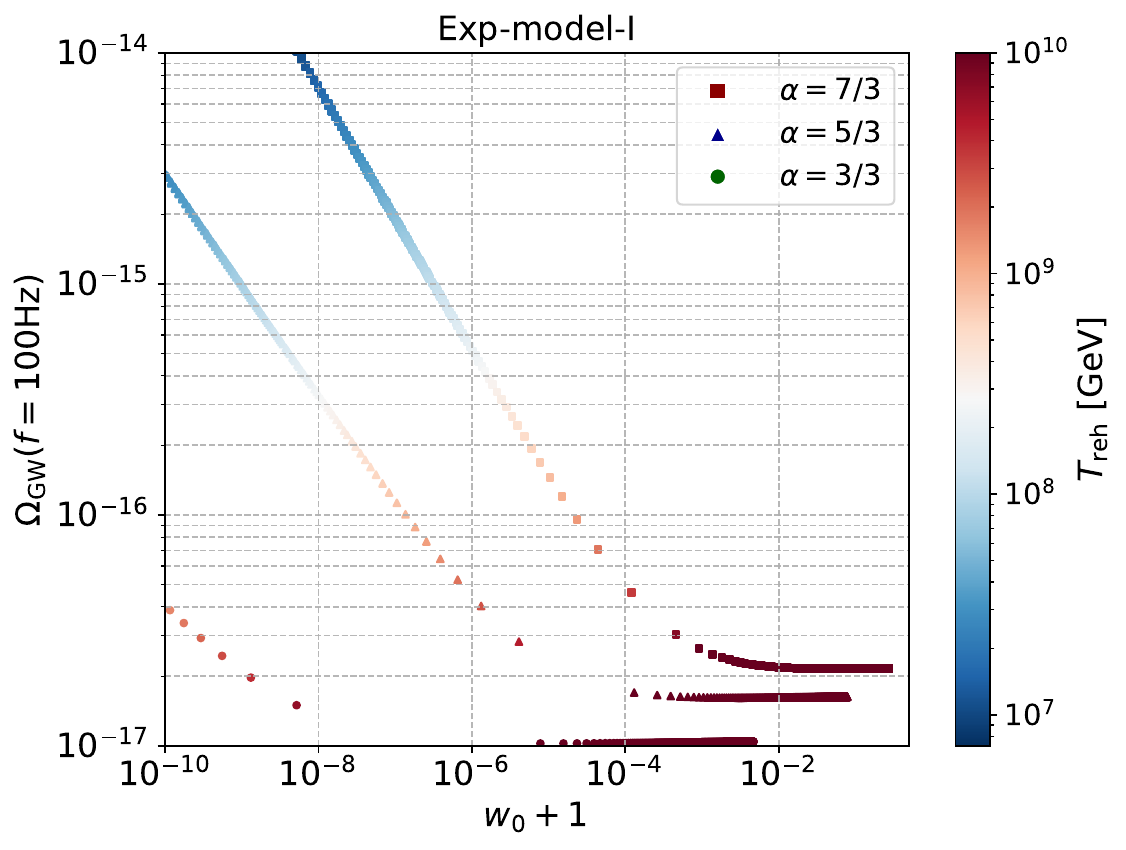}\\  
    \caption{Relationship between the value of the GW power spectrum $\Omega_{\rm GW}(f = 100\mathrm{Hz})$ and $w_0 + 1$ for Exp-model I, where $w_0$ denotes the present dark-energy equation-of-state parameter. Different colors of the points correspond to different values of $T_{\rm reh}$, while different shapes correspond to different values of $\alpha$.
    }
    \label{fig:w_Omega_GW_100I}
\end{figure}

\begin{figure}
\includegraphics[width=0.45\textwidth]{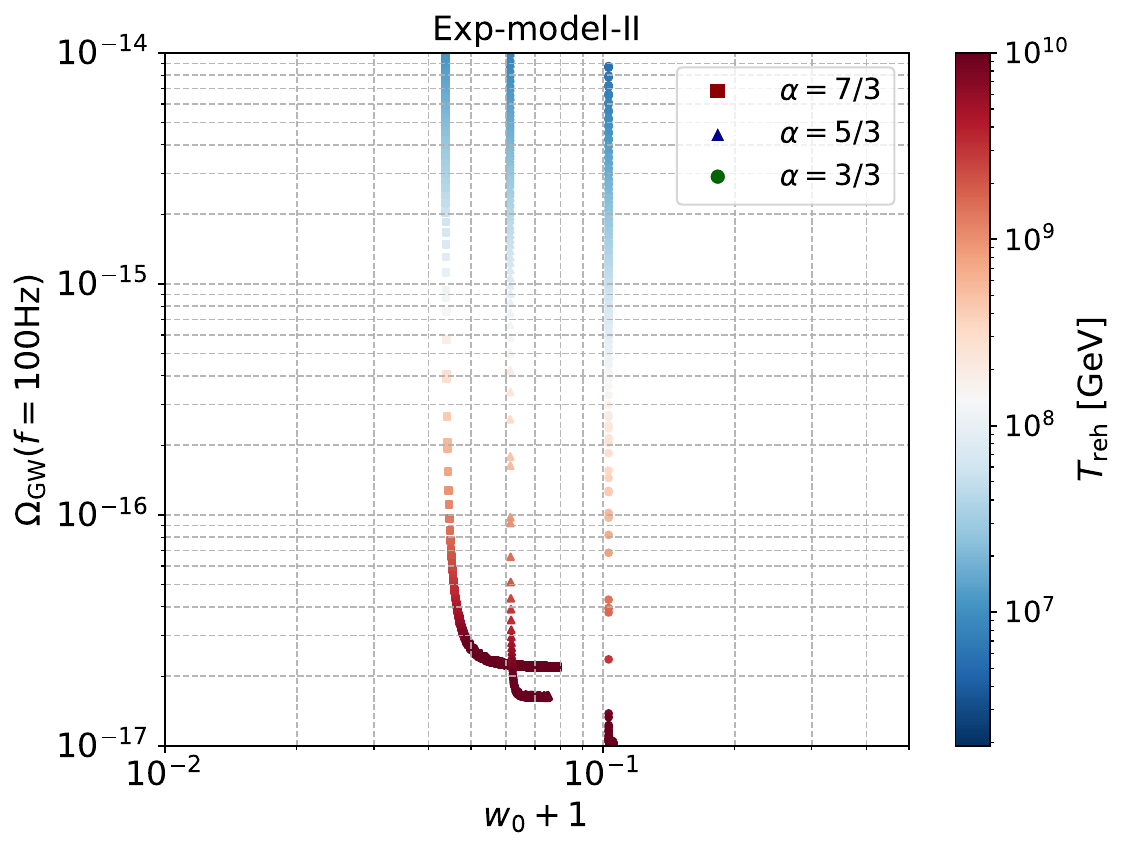}\\  
    \caption{Relationship between $\Omega_{\rm GW}(f = 100\mathrm{Hz})$ and $w_0 + 1$ for Exp-model II.}
    \label{fig:w_Omega_GW_100II}
\end{figure}

In Figs.~\ref{fig:w_Omega_GW_100I} and \ref{fig:w_Omega_GW_100II}, we show the relation between $\Omega_{\rm GW}(f=100{\rm Hz})$ and the present EOS parameter of dark energy $w_0$ for both models, considering $3\alpha \in \{3,5,7\}$. Here we also ensure $\ln(10^{10}A_s)$ is within the 1$\sigma$ confidence region of the P-ACT-LB constraints. Our results show that, for Exp-model-I, a high-frequency enhancement of the GW background and a dynamically evolving dark energy cannot be realized simultaneously, whereas in Exp-model-II these two features can coexist. 
Since our interest lies in dynamical dark energy, as favored by the DESI DR2 data~\cite{DESI2025_BAO_DR2}, and its potential correspondence with a high-frequency GW signal, Exp-model-I does not provide a phenomenologically interesting parameter space capable of realizing both features simultaneously. We therefore focus exclusively on Exp-model-II in the following.

\section{MCMC ANALYSIS}
\label{sec:MCMC data}
We perform a detailed MCMC analysis to constrain the parameter space of the $\alpha$-attractor quintessential model, focusing on Exp-model-II. This section summarizes the MCMC methodology and settings.

\subsection{Computation of the CMB and matter power spectra}

In order to compute CMB and galaxy-survey observables, we modify the CAMB code~\cite{Lewis_2000,Howlett_2012} by incorporating the $\alpha$-attractor model. Our code includes an external module that computes the full evolution of the scalar field, consistently tracking its dynamics from inflation and preheating through to the dynamical dark-energy phase. The input parameters of this module are $\alpha$, $M$, $g$, $h$, together with the standard cosmological parameters $\Omega_c h^2$, $\Omega_b h^2$, and $H_0$. We note that $\gamma$ is not treated as a free parameter; instead, for each set of input parameters we determine $\gamma$ via a shooting procedure that ensures the correct present-day value of $H_0$. The module then passes the information on the primordial power spectrum (namely the numerically computed values of $n_s$ and $A_s$) together with the freezing value of the scalar field $\varphi_{\rm f}$ to CAMB, which subsequently computes cosmological observables such as the CMB temperature anisotropy spectrum and the matter power spectrum relevant for galaxy surveys. We note that CAMB recomputes the scalar-field evolution by solving the corresponding EOM, thereby fully capturing the dynamical dark-energy evolution instead of adopting a $w_0$-$w_a$ parametrization.

In addition, the code includes a dedicated module for evaluating the GW power spectrum. This module follows the procedure described in Sec.~\ref{sec:GW_nu}. Based on the resulting GW power spectrum, we compute the effective number of neutrino species $N_{\nu}$ using Eq.~\eqref{eq:definenu}, 
which is then passed to CAMB so that its impact on the CMB power spectra is consistently included. 
Importantly, the module numerically resolves the GW power spectrum up to ultra-high frequencies, enabling a precise determination of $N_{\nu}$ without relying on order-of-magnitude estimates.


\subsection{Data}
The cosmological datasets used in our MCMC analysis consist of three different observations: BAO, CMB, and supernova observations. Specifically, we use the latest BAO measurements from DESI DR2~\cite{DESI2025_BAO_DR2} and the Pantheon+ Type~Ia supernova dataset~\cite{Riess:2021jrx}. 

For the CMB, we employ the CMB-only likelihood provided by the ACT collaboration~\cite{AtacamaCosmologyTelescope:2025blo}, which combines measurements from Planck~\cite{Planck:2018vyg} and ACT. Within this framework, we consider either Planck-only, ACT-only, or joint Planck+ACT TT/TE/EE power spectra. In the combined analysis, to avoid double counting arising from overlapping sky coverage, we include only Planck high-$\ell$ data for multipoles $\ell<1000$ in TT and $\ell<600$ in TE and EE. We use low-$\ell$ CMB data from the Planck Commander TT likelihood~\cite{Planck:2018vyg} and the Sroll2 EE likelihood~\cite{Planck2018_Sroll2EE}. Finally, we also include the ACT CMB lensing likelihood.

In this work, we consider two different data combinations. The first combines the joint Planck+ACT CMB dataset with ACT lensing, DESI DR2 BAO measurements, and Pantheon+ supernova data; we denote this combination as \texttt{P-ACT-LB2S}. The second combination uses the Planck CMB dataset alone (i.e.~without ACT), together with ACT lensing, DESI DR2, and Pantheon+, and is denoted as \texttt{P-LB2S}.

\subsection{MCMC setups}
We carry out Markov Chain Monte Carlo (MCMC) simulations using the public sampler Cobaya~\cite{Torrado_2021,ascl1910}.
For comparison, we carry out MCMC analyses for the following two models. The first is the flat $\Lambda$CDM model, characterized by five key parameters
$\{H_0,\, \Omega_c h^2,\, \Omega_b h^2,\, n_s,\, \mathcal{A}_s\}$.
The second is the $\alpha$-attractor model, which involves six key parameters
$\{H_0,\, \Omega_c h^2,\, \Omega_b h^2,\, \alpha,\, M/\sqrt{\alpha},\, \log_{10} g\}$,
while another reheating parameter is fixed to $h = 0.01$.

Note that because there is a strong degeneracy between $M$ and $\alpha$, which slows down MCMC sampling, 
we sample $\{\alpha,\, M/\sqrt{\alpha}\}$ instead of $\{\alpha,\, M\}$.

For both models, we adopt broad and physically motivated priors for all standard cosmological parameters.
For $\alpha$-attractor model and instant reheating, we choose $\alpha\in[2/3,10]$, $M/\sqrt{\alpha}\in[0.5,3]\times10^{-5}$, and $\log_{10}g\in[-5.5,-2]$. 

Finally, we evaluate the Bayesian evidence $\ln E_i$ for each model using the MCEvidence package~\cite{Heavens:2017afc}. We then report the logarithmic Bayes factors $\ln B$ of the $\alpha$-attractor model relative to $\Lambda$CDM for both the \texttt{P-ACT-LB2S} and \texttt{P-LB2S} datasets. We note that values of $|\ln B| > 5$ indicate strong evidence in favor of one model over the other.

\section{MCMC RESULTS AND DISCUSSION}
\label{sec:MCMC result}

\begin{table*}[t]
\centering
\caption{
Posterior means (best-fit values in parentheses) with $\pm1\sigma$ uncertainties for the $\alpha$-attractor Exp-model-II model and the $\Lambda$CDM model, inferred from the P--LB2S and P--ACT--LB2S datasets. Also reported are the minimum $\chi^2$ (overall and for each dataset), the Bayesian evidence $\ln E_i$, the log Bayes factor $\ln B$, and $\Delta\chi^2$ values.}
\label{tab:PLBS_PACTLBS_bestfit}
\begin{tabular}{lcccc}
\hline\hline
Model(Data) 
& $\alpha$-attractor(P--LB2S) 
& $\Lambda$CDM(P--LB2S)
& $\alpha$-attractor(P--ACT--LB2S) 
& $\Lambda$CDM(P--ACT--LB2S) \\
\hline
$H_0$                        
& $67.58(67.64)^{+0.45}_{-0.36}$ 
& $68.40(68.36)^{+0.2853}_{-0.2847}$
& $67.45(67.54)^{+0.37}_{-0.29}$ 
& $68.34(68.34)^{+0.26}_{-0.26}$ \\[3pt]

$\Omega_{\rm c}h^2$          
& $0.1180(0.1179)^{+0.00058}_{-0.00073}$ 
& $0.117617(0.1175)^{+0.000649}_{-0.000644}$
& $0.1183(0.1180)^{+0.00057}_{-0.00061}$ 
& $0.1175(0.1175)^{+0.00065}_{-0.00063}$ \\[3pt]

$\Omega_{\rm b}h^2$          
& $0.02253(0.02251)^{+0.00012}_{-0.00012}$ 
& $0.02253(0.02249)^{+0.0001193}_{-0.0001210}$
& $0.02260(0.02257)^{+0.00010}_{-0.00010}$ 
& $0.02256(0.02253)^{+0.00010}_{-0.00010}$ \\[3pt]

$\alpha$                     
& $4.236(2.7497)^{+1.157}_{-2.988}$ 
& ---
& $3.908(2.6048)^{+0.984}_{-2.404}$ 
& --- \\[3pt]

$\dfrac{M}{\sqrt{\alpha}}\times10^5(/M_{\rm Pl}^2)$   
& $1.047(1.028)^{+0.0178}_{-0.0170}$ 
& ---
& $1.044(1.027)^{+0.0156}_{-0.0155}$ 
& --- \\[3pt]

$\log_{10} g$                
& $-3.437~(-3.975)^{+0.6324}_{-0.6209}$ 
& ---
& $-3.443~(-4.040)^{+0.5087}_{-0.595}$ 
& --- \\[3pt]

$n_s$                        
& $0.9659~(0.9664)^{+0.00041}_{-0.00060}$ 
& $0.971112~(0.9712)^{+0.003316}_{-0.003292}$
& $0.9659~(0.9664)^{+0.00046}_{-0.00046}$ 
& $0.9747~(0.9752)^{+0.0030}_{-0.0030}$ \\[3pt]

$\ln 10^10 A_s$                   
& $3.061~(3.060)^{+0.0098}_{-0.011}$ 
& $3.062~(3.059)^{+0.010147}_{-0.011420}$
& $3.059~(3.058)^{+0.010}_{-0.011}$ 
& $3.061~(3.060)^{+0.010}_{-0.012}$ \\[3pt]

$\log_{10}(T_{\rm reh}/{\rm GeV})$       
& $8.173(7.712)^{+0.704}_{-1.291}$ 
& ---
& $8.233(7.661)^{+0.534}_{-1.125}$ 
& --- \\[3pt]
\hline
$\chi^2$
&2436.80
&2436.41
&2238.18
&2230.86\\
$\chi^2_{\rm DESI-DR2}$
&13.32
&11.85
&15.06
&12.35
\\
$\chi^2_{\rm Pantheon+}$
&1402.99
&1405.84
&1402.99
&1405.76
\\
$\chi^2_{\rm CMB}$
&1000.49
&999.09
&636.69
&633.75
\\
\hline
$\ln E_i$
&-2463.14
&-2458.38
&-2266.82
&-2254.35\\
$\ln B$
&-4.76
&---
&-12.47
&---\\
$\Delta\chi^2$
&0.39
&---
&7.32
&---
\\
\hline
\end{tabular}
\end{table*}





\begin{figure*}[h!]
    \includegraphics[width=0.9\textwidth]{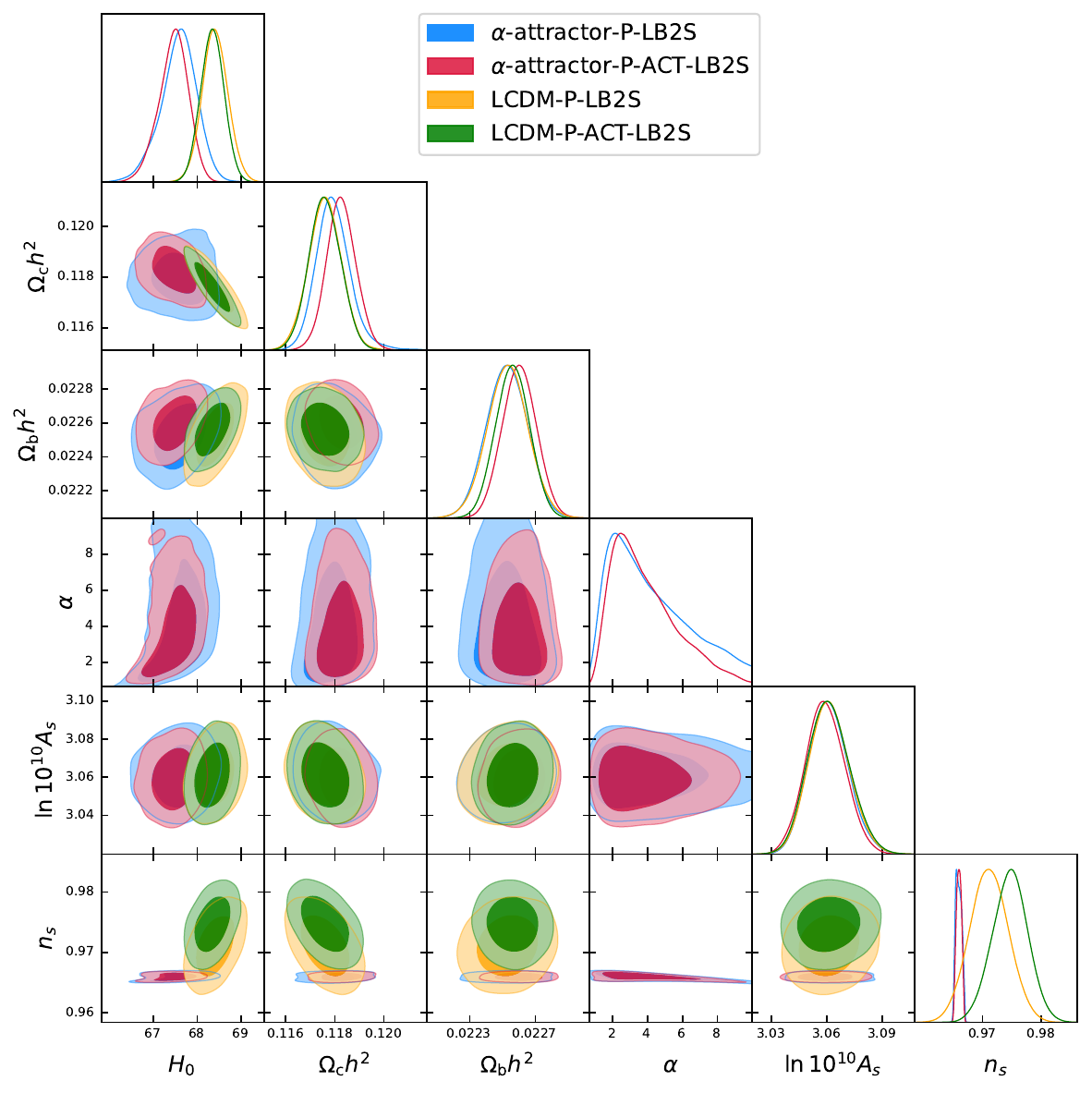}\\  
    \caption{
    The $63.3\%-95.5\%$ confidence contours for cosmological parameters of the $\alpha$-attractor Exp-model-II model (blue, red) and the $\Lambda$CDM model (yellow, green). Blue and yellow are derived from Planck 2018 CMB, ACT lensing, Pantheon+, and DESI DR2 (P-LB2S); red and green additionally include ACT CMB data (P-ACT-LB2S).}
    \label{fig:Full_triangle1}
\end{figure*}

\begin{figure}[h!]
    \includegraphics[width=0.45\textwidth]{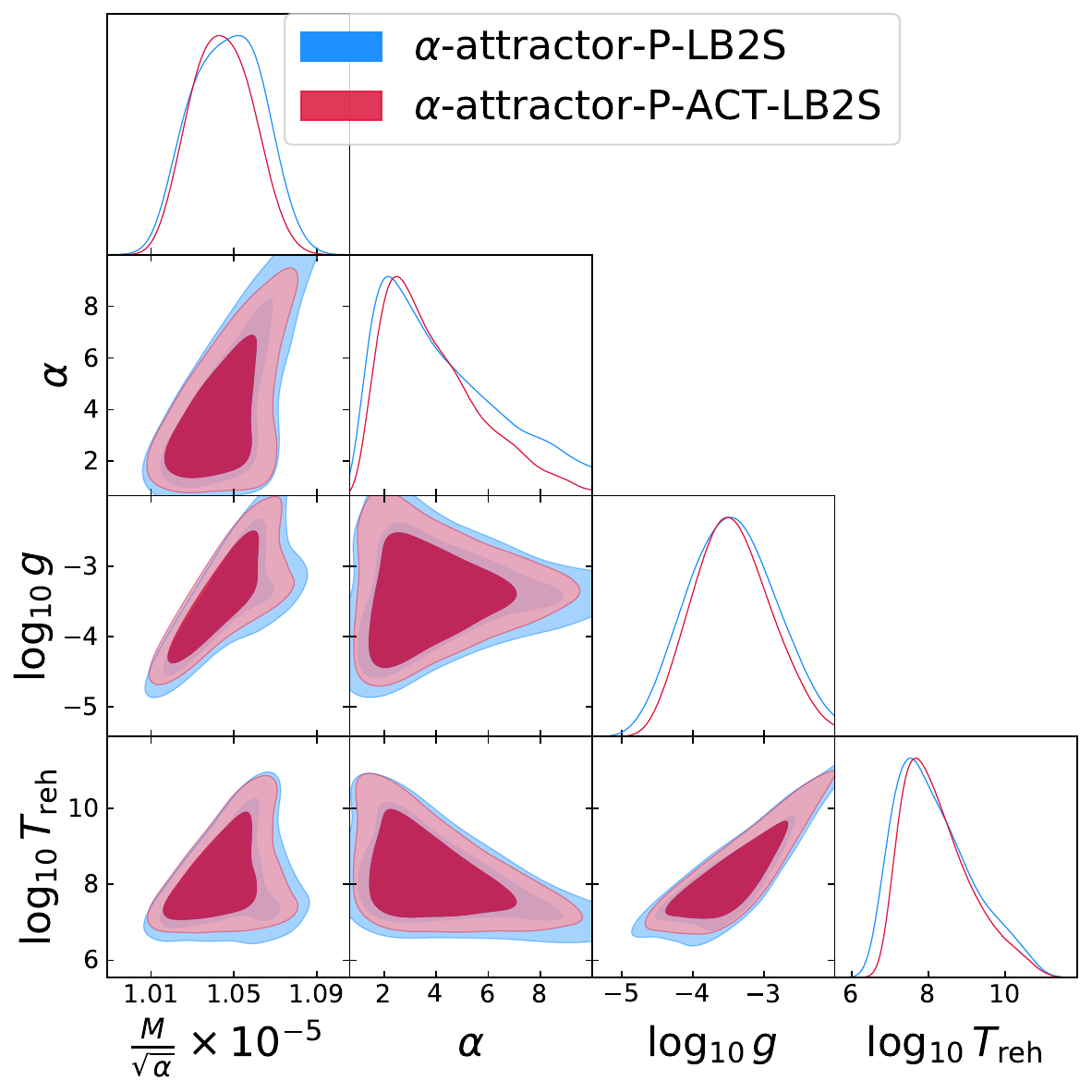}\\  
    \caption{
    The $63.3\%-95.5\%$ confidence contours for the model parameters of the $\alpha$-attractor Exp-model-II model. The blue contours use Planck 2018 CMB, ACT lensing, Pantheon+, and DESI DR2 (P-LB2S); the red contours additionally include ACT CMB data (P-ACT-LB2S).}
    \label{fig:Full_triangle2}
\end{figure}

\subsection{MCMC results}
Now, we present our MCMC results comparing the $\alpha$-attractor quintessential Exp-model-II model and the $\Lambda$CDM model for two data combinations: the \texttt{P-LB2S} and \texttt{P-ACT-LB2S} datasets. 
The main results of our MCMC analysis are presented in Table \ref{tab:PLBS_PACTLBS_bestfit} and shown in Figs. \ref{fig:Full_triangle1} and \ref{fig:Full_triangle2}. 

Table \ref{tab:PLBS_PACTLBS_bestfit} presents the mean values, best-fit points, and the corresponding 1-$\sigma$ marginalized uncertainties for the cosmological parameters of the $\alpha$-attractor model and the $\Lambda$CDM model, using both the \texttt{P-LB2S} and \texttt{P-ACT-LB2S} datasets. For the $\alpha$-attractor model, we show $n_s$, $\log A_s$, and $\log_{10}(T_{\rm reh}/{\rm GeV})$ as derived parameters. Finally, we also provide the minimal $\chi^2$, $\ln E_i$, $\ln B$, and $\Delta\chi^2$ for each model and data combination. 

In Fig.~\ref{fig:Full_triangle1}, we compare constraints on the $\alpha$-attractor model and the $\Lambda$CDM model, plotting the $63.3\%$ and $95.5\%$ confidence contours for the cosmological parameters $\{H_0,\, \Omega_c h^2,\, \Omega_b h^2,\, n_s,\, \mathcal{A}_s\}$. The $\alpha$-attractor model includes one additional parameter $\alpha$. In Fig.~\ref{fig:Full_triangle2}, we show constraints on the $\alpha$-attractor input parameters 
$\{ \alpha,\, M/\sqrt{\alpha},\, \log_{10} g\}$, which represent the model’s physical inputs. Observable quantities, $\{n_s, \mathcal{A}_s\}$ and the dark energy evolution, are determined numerically from these inputs, while $H_0$ (chosen at each MCMC sample) is used to fix the model’s additional parameter $\gamma$. We also show the posterior distribution of the reheating temperature $\log_{10} (T_{\rm reh}/{\rm GeV})$, which is derived from the input parameters and is strongly correlated with the coupling constant $g$. It is not completely degenerate with $g$ because it also depends on the value of $\dot{\phi}$ at $\phi=0$, which in turn depends on the shape of the inflationary potential, i.e., on $M$ and $\alpha$.  

Notably, the reheating temperature is tightly constrained as we find $ T_{\rm reh} \in \left[6.25\times10^{6},\,1.35\times10^{10}\right]\ \mathrm{GeV}$ at the 95\% confidence level for \texttt{P-ACT-LB2S}. The lower bound on $\log_{10} (T_{\rm reh}/{\rm GeV})$ is set by the GW constraint via $\Delta N_{\rm eff}$, since lower reheating temperatures raise the GW peak and can lead to GW overproduction as discussed in Sec. \ref{sec:GW_nu}.  On the other hand, the upper bound comes from the lower limit on $n_s$ through Eqs.\eqref{eq:n_s_alpha} and \eqref{eq:N*}. Moreover, $T_{\rm reh}$ also affects the late-time evolution of dark energy by determining the value of $\phi_{\rm f}$, as discussed in Sec.~\ref{sec:dynamic-dark-energy}. Excessively large $T_{\rm reh}$ is disfavored because it leads to overly dynamical dark-energy behavior that is incompatible with observational constraints, providing an additional upper bound on $T_{\rm reh}$.

The best-fit value of $\alpha$ from the \texttt{P-LB2S} dataset is $\alpha=2.75$. When ACT data are included, the preferred value shifts slightly lower to $\alpha=2.60$. In both cases, the posterior for $\alpha$ exhibits a tail toward large values. This arises because, as $\alpha$ increases, the dark-energy evolution becomes increasingly insensitive to $\alpha$ (see Fig. 17 of Ref.\cite{Akrami:2017cir}), making the model effectively degenerate in the large-$\alpha$ regime.

As shown in Table \ref{tab:PLBS_PACTLBS_bestfit}, we find $\ln B=-4.76$ with \texttt{P-LB2S}, and $\ln B=-12.47$ with \texttt{P-ACT-LB2S}, implying a strong preference for $\Lambda$CDM over the $\alpha$-attractor model. This result is significantly different from Ref.~\cite{alestas_desi_2025}.
This is primarily because we include the effect of $\Delta N_{\rm eff}$ in the analysis. As discussed in Sec.~\ref{sec:GW_nu}, GW constraints require the scalar spectral index $n_s$ to be smaller than the best-fit value obtained when the $\Delta N_{\rm eff}$ effect is neglected. The tension increases when ACT data are included, since ACT tends to prefer higher $n_s$ values.

\subsection{Gravitational waves}
\begin{figure}
    \centering
    \includegraphics[width=0.45\textwidth]{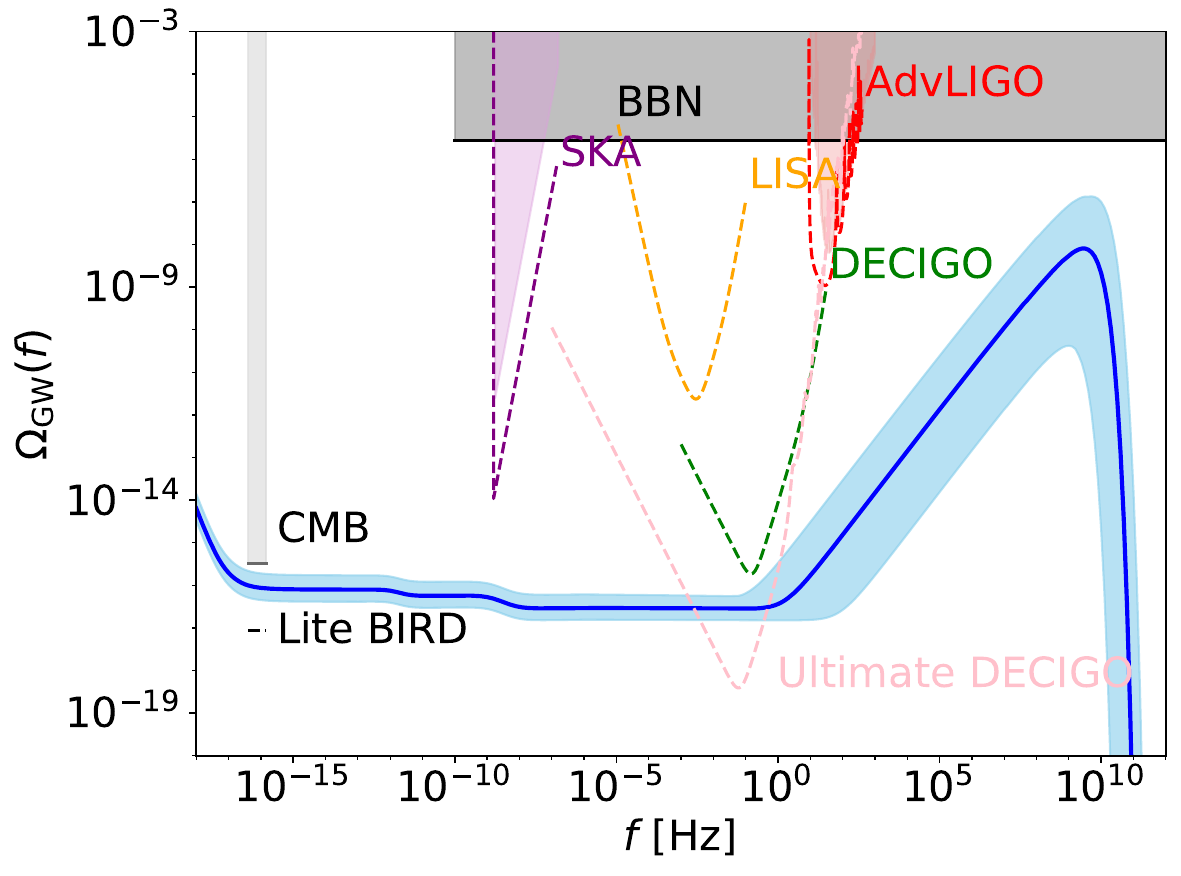}
    \caption{GW power spectrum corresponding to the best-fit $\alpha$-attractor Exp-model-II in Table \ref{tab:PLBS_PACTLBS_bestfit} for P-ACT-LB2S, together with the sensitivities of current (solid lines with gray shading) and future (dashed lines) GW experiments. The blue shaded band shows the corresponding $1\sigma$ region.
}
    \label{fig:GW_1sigma}
\end{figure}

In Fig.~\ref{fig:GW_1sigma}, we show the predicted GW power spectrum for the $\alpha$-attractor Exp-model-II model, plotted for the best-fit parameters (blue solid line) together with the $68.3\%$ confidence region of the parameter space (blue shaded region).

Thanks to the DESI data, we obtain tighter constraints on $\alpha$, establishing a stronger lower bound that in turn implies a lower limit on the tensor-to-scalar ratio and underscores the model’s predictive power for the amplitude of primordial GWs. 
This lower bound suggests the primordial GW background of this model could be tested by next-generation CMB B-mode experiments such as LiteBIRD~\cite{LiteBIRD:2022cnt}.

Furthermore, the CMB constraint on $\Delta N_{\rm eff}$ places a lower bound on the reheating temperature $T_{\rm reh}$, which further sharpens predictions for the GW amplitude at high frequencies. Our best-fit model indicates the primordial GW power spectrum shows a noticeable enhancement at frequencies above $\sim 1\,\mathrm{Hz}$. The predicted signal from the $\alpha$-attractor model remains too weak for current or near-future ground- and space-based laser-interferometer detectors. However, ultimate DECIGO sensitivity would be sufficient to detect it \cite{kudoh_detecting_2006,Kuroyanagi:2014qza}, and for some parameter values allowed by our constraints the reheating frequency $f_{\rm reh}$ falls within DECIGO’s detectable range. This raises the prospect of probing the reheating temperature via future GW observations~\cite{Kuroyanagi:2011fy}.


The peak of the primordial GW spectrum lies near $10^{10}\mathrm{Hz}$, well beyond the reach of current or planned laser-interferometer detectors but potentially accessible to future resonant-cavity experiments \cite{PhysRevD.105.116011}.

\subsection{Dynamical dark energy\label{sec:dde}}

\begin{figure}
    \centering
    \includegraphics[width=0.45\textwidth]{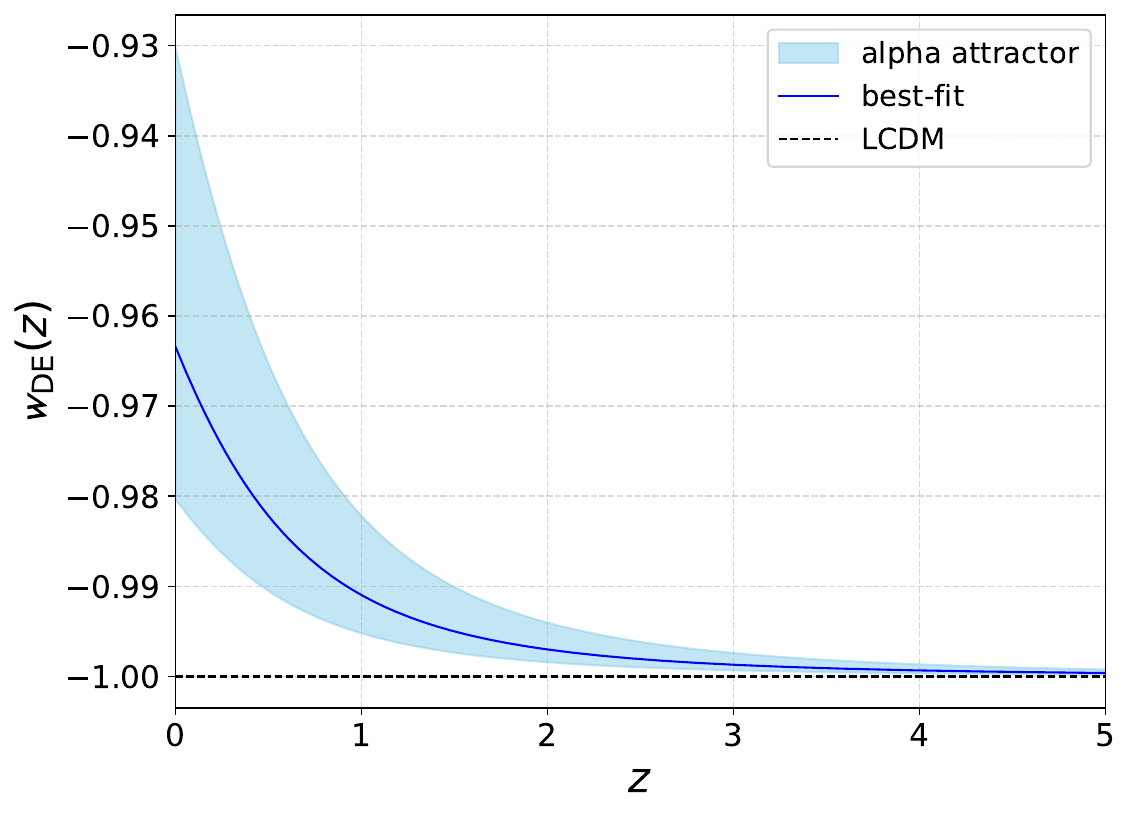}
    \caption{The dark-energy EOS parameter $w_{\rm DE}(z)$ as a function of redshift $z$ for the best-fit $\alpha$-attractor Exp-model-II (blue solid line), with the corresponding $1\sigma$ confidence region (blue shaded band).} 
    \label{fig:wz_1sigma}
\end{figure}

\begin{figure}
    \centering
    \includegraphics[width=0.45\textwidth]{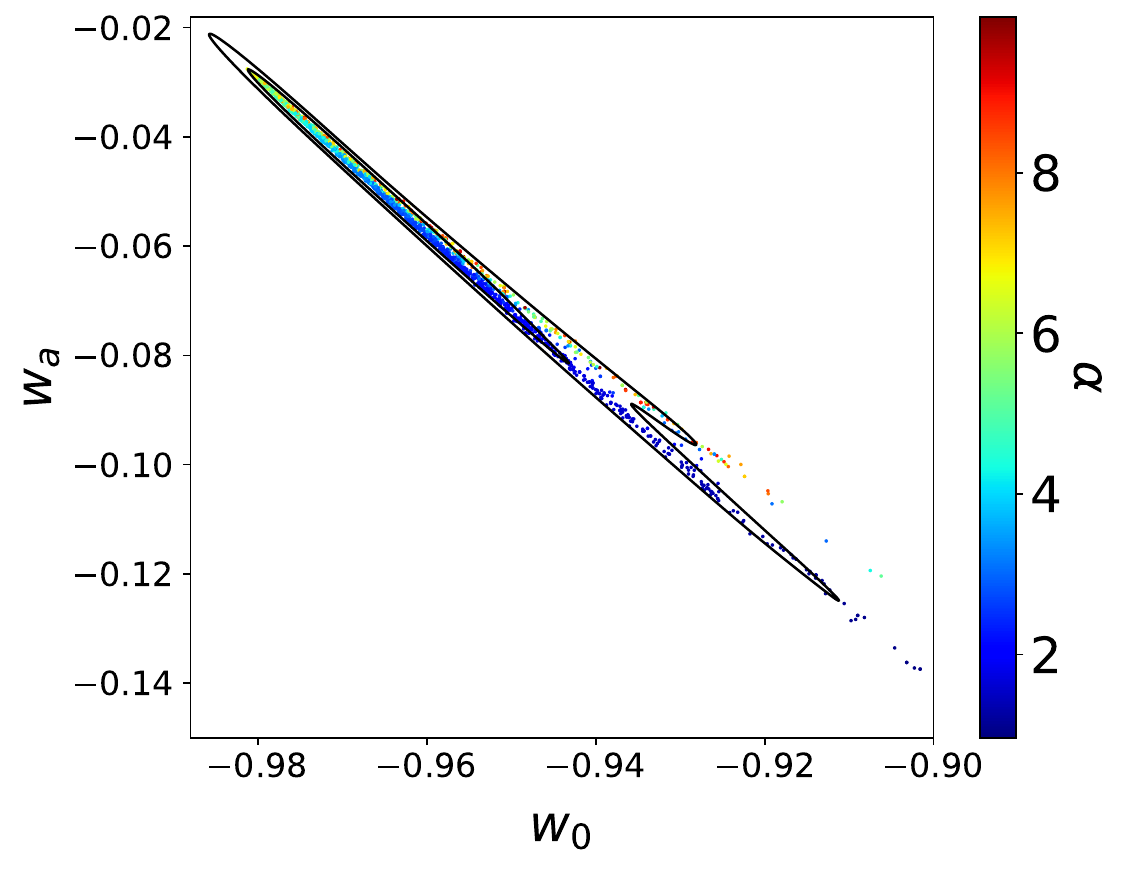}
     \caption{The $68.3\%$ and $95.5\%$ confidence contours for the present-day dark energy EOS parameter $w_0$ and its time derivative $w_a$, constrained by the P-ACT-LB2S data set. The color bar indicates the corresponding values of $\alpha$.}
     \label{fig:CPL_sigma}
\end{figure}

In Fig.~\ref{fig:wz_1sigma}, we show the dark-energy EOS parameter $w_{\rm DE}(z)$ as a function of redshift $z$ for the best-fit $\alpha$-attractor Exp-model-II model (blue solid line). The blue shaded region denotes the $68.3\%$ confidence region of the parameter space. In Fig.~\ref{fig:CPL_sigma}, we present the constraints in terms of the present-day dark-energy EOS parameter $w_0$ and its time variation $w_a$. Both plots indicate that the data prefer dynamical dark energy. 

In Fig.~\ref{fig:CPL_sigma}, we see that the $2\sigma$ confidence region splits into two distinct branches in the lower-right corner, where dark energy becomes more dynamical, characterized by larger $w_0$ and smaller $w_a$. These two branches correspond to different pathways to dynamical dark energy. For the lower branch, the $\alpha$-attractor field is more dynamical due to smaller values of $\alpha$. For the upper branch, the models have higher reheating temperatures, causing the scalar field to freeze earlier in a steeper part of the potential. As a result, when the field resumes rolling at late times, the dark-energy dynamics become more pronounced.




\section{Conclusions \label{sec:conclusions}}
In this work, we studied in detail the 
$\alpha$-attractor quintessential inflation model with instant reheating and 
its associated GW signal.
Using the latest cosmological data, including CMB temperature and polarization spectra and CMB lensing from Planck 2018 and ACT DR6, the latest DESI BAO measurements, and the Pantheon+ Type Ia supernova sample, we constrained the 
model parameters. 

The novel aspect of this study is that we included the contribution of primordial GWs to the effective number of relativistic degrees of freedom, thereby constraining a new region of the model’s parameter space. For this purpose, we performed full numerical computations that solve the full evolution of the quintessential scalar field and incorporate detailed reheating dynamics. This enables us to evaluate the GW spectral amplitude at the high-frequency peak and to accurately compute the evolution of dynamical dark energy.

By calculating the Bayes factor, we compared the $\alpha$-attractor quintessential model with $\Lambda$CDM and found that the $\alpha$-attractor model is disfavored by the data. This is driven by a tension in $n_s$. Since the constraint on $\Delta N_{\rm eff}$ does not allow an excessively low reheating temperature $T_{\rm reh}$, the lower bound on $T_{\rm reh}$ translates into an upper bound on the e-folding number $N_*$, and consequently into an upper bound on the scalar spectral index $n_s$. However, the value of $n_s$ favored by CMB observations tends to be larger than this upper bound from GW overproduction, a discrepancy that becomes particularly pronounced when ACT data are included. 

Although the model is disfavored, we also provided predictions for the primordial GW power spectrum, comparing it with the sensitivities of future GW experiments. Thanks to the tightly constrained model parameters, we obtain highly predictive forecasts for the GW spectrum.

Finally, let us discuss the possible generalization of our results to other reheating mechanisms. In the case of gravitational reheating, the reheating temperature is typically low, $T_{\rm reh}<10^6$GeV. This prolonged kination era leads to an overproduction of GWs, which violates observational bounds. Consequently, gravitational reheating is clearly disfavored in this context.

The discussion is somewhat more complicated for reheating mechanisms predicting higher reheating temperatures, because they evade GW constraints and the behavior of dynamical dark energy plays an important role in constraining the model. Especially when the backreaction effect is important, the freeze‑out value of the scalar field $\varphi_{\rm f}$, that affects the behavior of the dynamical dark energy component, becomes model-dependent. On the other hand, in the regime where the reheating mechanism does not affect the dynamics of the scalar field, the situation remains simple. In this case, $\varphi_{\rm f}$ is determined solely by Hubble friction due to the radiation background and is fixed by the amount of radiation produced, which is characterized by the reheating temperature $T_{\rm reh}$. In other words, there is a one‑to‑one correspondence between $\varphi_{\rm f}$ and $T_{\rm reh}$, regardless of the reheating model. In the instant preheating scenario, backreaction is negligible for $g<10^{-3}$~\cite{PhysRevD.97.063525}, which is the region where most of our 2$\sigma$ constraint lies. Therefore, within this parameter range, our results apply generically to other reheating mechanisms, under the assumption that they do not affect the scalar field evolution. As an example, this includes curvature reheating scenarios in which no direct interaction with the inflaton is present \cite{dimopoulos_non-minimal_2018}, so that there is no backreaction; in this case, the reheating process influences $\varphi_{\rm f}$ only through Hubble friction.

To conclude, our analysis highlights the constraining power of the diverse cosmological datasets, which is further strengthened by the inclusion of GW constraints. As forthcoming GW probes are developed, including CMB polarization experiments and interferometers covering a wide range of frequency bands, we expect to place much tighter constraints on a broad class of inflationary models. 


\section*{Acknowledgements}
The authors acknowledge the use of the Finis Terrae III supercomputer, which is part of the Centro de Supercomputacion de Galicia (CESGA) and is funded by the Ministry of Science and Innovation, Xunta de Galicia and ERDF (European Regional Development Fund). The authors also acknowledge support from the Spanish Research Agency (Agencia Estatal de Investigaci\'on) through the Grant IFT Centro de Excelencia Severo Ochoa No CEX2020-001007-S, funded by MCIN/AEI/10.13039/501100011033. SK is supported by the I+D grant PID2023-149018NB-C42 funded by MCIN/AEI/10.13039/501100011033, the Leonardo Grant for Scientific Research and Cultural Creation 2024 from the BBVA Foundation, and Japan Society for JSPS KAKENHI Grant no. JP23H00110 and JP24K00624.
CCJ is supported by the China Scholarship Council (Grant No.202406040043). G.A. is supported by the Spanish Research Agency’s Consolidaci\'on Investigadora 2024 grant CNS2024-154430. 

\bibliographystyle{apsrev4-1}
\bibliography{Bibliography}

\end{document}